\documentclass[fleqn,usenatbib]{mnras}

\pdfoutput=1
\usepackage{newtxtext,newtxmath}

\usepackage[T1]{fontenc}

\DeclareRobustCommand{\VAN}[3]{#2}
\let\VANthebibliography\thebibliography
\def\thebibliography{\DeclareRobustCommand{\VAN}[3]{##3}\VANthebibliography}

\usepackage{graphicx}
\usepackage{dcolumn}
\usepackage{bm}
\usepackage{hyperref}
\usepackage[mathlines]{lineno}
\usepackage{amsmath,euscript}
\usepackage{calrsfs}

\usepackage{tikz}
\usetikzlibrary{calc,decorations.markings}
\usetikzlibrary{snakes}
\usepackage{rotating}
\usepackage{comment}
\usepackage{empheq}
\usepackage{ulem}

\newcommand{\abs}[1]{\left|#1\right|}

\newcommand{\CE}{\mathcal{E}}
\newcommand{\CB}{\mathcal{B}}

\newcommand{\CO}{\mathcal{O}}

\newcommand{\spa}{\ , \ \ }

\allowdisplaybreaks[2]
\numberwithin{equation}{section}

\definecolor{darkgreen}{rgb}{0.0, 0.55, 0.1}

\newcommand{\lbreak}{\\[-2mm]}

\title[Binary mergers in strong gravity background of Kerr black hole]{Binary mergers in strong gravity background of Kerr black hole}

\author[Filippo Camilloni et al.]{
Filippo Camilloni,$^{3}$\thanks{camilloni@itp.uni-frankfurt.de }
Troels Harmark,$^{2}$\thanks{harmark@nbi.ku.dk}
Gianluca Grignani$^{1}$\thanks{gianluca.grignani@unipg.it}
Marta Orselli$^{1,2}$\thanks{orselli@pg.infn.it}
and Daniele Pica$^{1,2}$\thanks{daniele.pica@nbi.ku.dk}
\\
$^{1}$Dipartimento di Fisica e Geologia, Universit\`a di Perugia, I.N.F.N. Sezione di Perugia, Via Pascoli, I-06123 Perugia, Italy\\
$^{2}$Niels Bohr International Academy, Niels Bohr Institute, Copenhagen University, Blegdamsvej 17, DK-2100 Copenhagen \O{}, Denmark\\
$^{3}$Institut f\"ur Theoretische Physik, Goethe Universit\"at, Max-von-Laue-Str. 1, 60438 Frankfurt am Main, Germany\\
(Authors appear in alphabetical order)
}

\date{Accepted 2024 April 11. Received 2024 february 28; in original form 2023 November 22}

\pubyear{2024}

\begin{document}
\label{firstpage}
\pagerange{\pageref{firstpage}--\pageref{lastpage}}
\maketitle

\begin{abstract}
Binary-black-hole (BBH) mergers can take place close to a supermassive black hole (SMBH) while being in a bound orbit around the SMBH. In this paper, we study such bound triple systems and show that including the strong gravity effects of describing the SMBH with a Kerr metric can significantly modify the dynamics, as compared to a Newtonian point particle description of the SMBH. 
We extract the dynamics of the system, using a quadrupole approximation to the tidal forces due to the SMBH. We exhibit how the gyroscope precession is built into this dynamics, and find the secular Hamiltonian by both averaging over the inner and outer orbits, the latter being the orbit of the BBH around the SMBH. We study the long-time-scale dynamics, including the periastron precession and GW radiation-reaction of the binary system, finding that the strong gravity effects of the SMBH can enhance the von Zeipel-Lidov-Kozai  mechanism, resulting in more cycles, higher maximum eccentricity, and thereby a shorter merger time, particularly when the binary is close to, or at, the innermost stable orbit of the SMBH. We end with an analysis of the peak frequency of the GW emission from the binary system, highlighting possible observable signatures in the ET and LISA frequency bands.
\end{abstract}




\section{Introduction} 
\label{sec:intro} 

Gravitational waves (GWs) from merging binary black holes (BBHs) - as well as other compact objects - recently opened up a new window for observations of gravitational physics \cite{LIGOScientific:2016aoc, LIGOScientific:2020ufj,LIGOScientific:2021qlt}. 
While observations are still in their infancy, it is generally believed that one has a significant number of bound systems of BBHs in the vicinity of a given supermassive black hole (SMBH), {\sl i.e.} at the galactic center (GC)~\cite{Antonini:2012ad,Generozov:2018niv,Fragione:2018yrb,Tagawa:2019osr}.  
This makes it crucial to investigate the signatures of GWs from such BBHs. 
In particular, it has been shown that the vicinity of the SMBH can give rise to shorter merger times through the von Zeipel-Lidov-Kozai (ZLK)~\cite{1910AN....183..345V, 1976CeMec..13..471L,doi:10.1146/annurev-astro-081915-023315} mechanism, along with a high eccentricity ($e > 0.1$) when entering in the LIGO-Virgo-KAGRA window of the merger \cite{Samsing:2017xmd,Samsing:2017rat,Randall:2018qna,Samsing:2020tda,Tagawa:2020jnc,Trani:2021tan,Rom:2023kqm}.
\lbreak

So far, nearly all of the literature has studied BBHs in vicinity of a SMBH
using the approximation that all three BHs are particles.
This is valid provided the BHs only affect each other through weak gravity and with relative velocities much smaller than the speed of light, such that the effects due to General Relativity (GR) can be added through Post-Newtonian (PN) corrections.  
\lbreak

However, some BBH mergers can conceivably take place very close to the SMBH, perhaps even near its innermost stable circular orbit (ISCO) \cite{Peng_2021}.
In this case the particle/PN approximation of the supermassive black hole breaks down since it is only valid if the distance between the BBH and the SMBH is much larger than the Schwarzschild radius of the SMBH. The consequence of this breakdown is that one can only study the system using a metric for the supermassive BH. Indeed, the BBH is exposed to strong gravity effects since its orbit around the SMBH can have a relative velocity comparable to the speed of light and have effects of the gravitational potential that can be considered non-perturbative in comparison with the PN expansion. 
\lbreak

In this paper we study a BBH exposed to such strong gravity effects with the SMBH described via the Kerr metric.
\footnote{As we comment further on below, \cite{Maeda:2023tao} studied a Newtonian binary system in circular orbit around a large Schwarzschild BH along with results on ZLK oscillations. Our work differs in several ways, but mainly that we consider a large Kerr BH and we include post-Newtonian effects in the binary system such as periastron precession and radiation-reaction effects which have a major effect at long time-scales. Note also footnote \ref{foot8} regarding how the gyroscope precession enters in the secular Hamiltonian.}
The only necessary assumption is that the size of the binary system is very small compared to the curvature scale of the SMBH. This alone means that we can model the interaction with the SMBH through tidal forces. We use the quadrupole approximation but one can extend our study with further tidal correction terms. Moreover, in all cases that we study, we take the masses in the binary system to be much smaller than that of the SMBH, which means we can study the BBH-SMBH system even when the BBH is near or at the ISCO of the SMBH. 
\lbreak

As a further simplification, we assume that the distance between the BHs in the binary system is sufficiently large, compared to both their Schwarzschild radii, so that one can use the particle/PN approximation for the BBH. This requirement significantly simplifies our analysis and it means we fully focus on new effects arising from treating the SMBH via a curved space-time metric instead of as a particle. 
\lbreak

Recently hierarchical three body systems have been studied in different papers,   using a two-body description to analyse the problem \cite{PhysRevD.104.024016}, including also 1PN and 1.5 PN effects \cite{Kuntz:2022onu,PhysRevD.99.103005}, and the context of gravitational waves \cite{Fang:2019mui,PhysRevD.102.104002} even in the strong field regime \cite{Cardoso:2021vjq}.
\lbreak

The main goal of this paper is to study the long-time-scale dynamics of the BBH-SMBH system, and to understand how this is modified by strong gravity effects arising from a metric description of the SMBH, as opposed to a Newtonian point particle description. 
\lbreak

To accomplish this goal, we use the first part of the paper (Secs.~\ref{sec:binary_general} to \ref{sec: <H>}) to set up and derive the dynamical equations of the BBH-SMBH system, ending with the secular Hamiltonian that describes the long-time-scale dynamics. 
In the second part (Secs.~\ref{sec: relativistic_effects_KL} and \ref{sec: GW}) we use the secular Hamiltonian to study the long-time-scale dynamics, including the interplay between the ZLK mechanism, the periastron precession, and the GW radiation-reaction forces in the binary system. As we shall see, the strong gravity effects of the SMBH can significantly change the dynamics of the binary system when it is close to the SMBH.
\lbreak

We begin the first part of the paper, in Secs.~\ref{sec:binary_general} and \ref{sec:binary_kerr}, with deriving a Hamiltonian for a BBH system in the vicinity of an SMBH described by the Kerr metric. The center of mass motion of the BBH is in our approximation moving on a geodesic of the Kerr metric. For simplicity, we consider a circular equatorial geodesic. The effect of the SMBH on the BBH is described via quadrupole tidal forces, using the work of Marck \cite{Marck:1983,Camilloni:2023rra}. At this point, we take the Newtonian approximation of the internal dynamics of the binary system, while the tidal forces are the full GR result from the Kerr metric, which includes strong gravity effects. 
\lbreak

In Sec.~\ref{sec:gyroscope} we discuss the gyroscope precession effect on the SMBH-BBH system. This warrants changing the frame of reference, from a local inertial frame of reference, as found by Marck in \cite{Marck:1983}, to a non-inertial frame of reference that we call {\sl distant-star} frame as found in \cite{Bini:2016iym}. The distant-star frame provides a global point of view on the orientation of the SMBH-BBH system since the frame orientation can be seen as aligning with the orientation of an asymptotic observer. In this frame, the gyroscope precession angle is directly built into the local dynamics. 
\lbreak

In Sec.~\ref{sec: euler_angles_and_action_angle_variables} we exhibit the appropriate Euler angles as well as action-angle variables in the distant-star frame, thereby setting up our study of the dynamics.
\lbreak

In Sec.~\ref{sec: <H>} we reach the goal of the first part of the paper, with the derivation of the long-time-scale dynamics of the BBH-SMBH system including the strong gravity effects arising from a metric description of the SMBH
In detail, we find the secular Hamiltonian in Sec.~\ref{sec: <H>} by averaging both over the inner orbit of the BBH, as well as the outer orbit, the latter corresponding to the orbit of the BBH around the SMBH.   
An important consistency check on the secular Hamiltonian is that we obtain the same equivalent dynamics from both Marck's frame of reference as well as the distant-star frame of reference. This is non-trivial as the outer orbit average is taken over different period angles. As we explain, this is all tied together by the gyroscope precession. 
\lbreak

We begin the second part of the paper with  Sec.~\ref{sec: relativistic_effects_KL} in which we take the first steps towards understanding the long-time-scale dynamics. This type of dynamics is interesting due to the ZLK mechanism \cite{1976CeMec..13..471L,1962AJ.....67..591K, doi:10.1146/annurev-astro-081915-023315} which is an astrophysically relevant effect observed in triple systems consisting of an exchange between the angular momentum of the inner and outer binary which has an impact on the eccentricity and inclination angle of the inner binary.~\footnote{In this paper ``inner" binary refers to the BBH system, while the ``outer" binary is given by the BBH-SMBH system.}
We show that a change in the ZLK frequency $\Omega_{\mbox{\tiny ZLK}}$ is the sole difference between the purely Newtonian triple system and our triple BBH-SMBH system in which we consider the full metric of the SMBH. 
Therefore, we make a careful study of $\Omega_{\mbox{\tiny ZLK}}$, including both the frequency $\Omega_{\mbox{\tiny ZLK}}^{\rm (GR)}$ that one obtains using the local proper time $\tau$ of the BBH, but also the frequency $\Omega^{(\infty)}_{\mbox{\tiny ZLK}}$ measured by a faraway observer, the difference being a redshift factor. Both of these frequencies are relevant to understand the dynamics of the BBH.
We compare these frequencies to $\Omega_{\mbox{\tiny ZLK}}^{\rm (N)}$ being the frequency obtained by describing all three black holes as Newtonian point particles. 
In particular, we find the important result that 
$\Omega_{\mbox{\tiny ZLK}}^{\rm (GR)}$ is twice $\Omega_{\mbox{\tiny ZLK}}^{\rm (N)}$ when the binary system moves on the ISCO, no matter what the SMBH spin is.  This is a key to understand many of the further results of this paper.
\lbreak

In Sec.~\ref{sec: relativistic_effects_KL} we also consider the large radius limit of the outer orbit, thus making contact with previous results in the literature for which one studied PN corrections to the long-time-scale dynamics of the triple system \cite{Liu_2019}. We, furthermore, predict new higher-order terms.
\lbreak

The paper culminates with Sec.~\ref{sec: GW}, in which we add the PN corrections of periastron precession and radiation-reaction effect of GW emission to the long-time-scale evolution equations of the binary system, following a similar procedure as \cite{Randall:2018qna}, but now including the effect of using the Kerr metric to describe the SMBH. 
The added PN corrections are highly important to the dynamics of the binary system. The periastron precession is important as it can counteract the ZLK mechanism, narrowing the window of possible initial inclination angles for which it can occur. Furthermore, the radiation-reaction effect of GW emission from the binary system models how the binary system loses angular momentum and energy, ending with the binary merger. 
\lbreak

The main result of Sec.~\ref{sec: GW} is that the strong gravity effect, arising from treating the SMBH via a metric, significantly changes the dynamics of the binary system: 
\begin{itemize}
\item
The strong gravity effects are enhanced by a closer proximity of the BBH to the SMBH. 
We study in particular when the BBH is moving along the ISCO of the SMBH, both with and without spin, corresponding to the largest possible strong gravity effects of the SMBH in our model. 
\item 
The strong gravity effects can enhance the ZLK mechanism, resulting in more ZLK cycles, a larger peak eccentricity, and thereby a shorter merger time, since more energy is emitted by GWs when the  eccentricity is high. 
    \item 
The spin of the SMBH can significantly impact the dynamics of the binary system. 
We find that it plays a significant role in whether the SMBH has zero spin, or whether it counter-rotates or co-rotates with respect to the outer orbit of the binary system around the SMBH. 
\item
Another important aspect of the binary dynamics is the redshift factor, which is used to translate the local time scales of the binary system to the asymptotic time scales as measured by a far away observer. The redshift factor is always greater than 1, increases with counter-rotating spin and decreases with co-rotating spin.
\item 
The various effects have an interesting interplay, with the ZLK mechanism enhanced by strong gravity, accelerating the merger of the binary system, while the redshift factor in a sense slows it down, since the local clock near the binary goes slower than that of an asymptotic observer. 
\end{itemize}
To look for potentially observable signatures for the third-generation interferometer Einstein Telescope (ET), as well as for the Laser Interferometer Space Antenna (LISA), we examine the peak frequency $f_{\rm GW}$ of the spectrum of GWs emitted from the binary system.
We find that it is conceivable that one can observe the ZLK cycles of the pre-merger phase of the binary system. Compared to a Newtonian point particle description of the SMBH, the strong gravity effects can result in a larger number of ZLK cycles as well as shorter merger times. Moreover, the maximum peak frequency is lowered by the redshift factor. 
\lbreak

We end the paper with our conclusions and give an outlook on future directions in Sec.~\ref{sec: concl}.
\lbreak

{\sl Note added: While the preparation of this paper was in its final stages, the preprint of \cite{Maeda:2023uyx} appeared with some overlap to this work.}

\section{Binary system in a tidal force background}
\label{sec:binary_general}

In this section, we review the basic dynamics of a binary system of black holes in a general tidal force background up to the quadrupole approximation, where the tidal forces originate from the curvature of the space-time that the binary system moves in. For simplicity, we take the Newtonian approximation of the black holes for which we can regard them as particles. Instead, the quadrupole tidal force background is kept completely general. We assume the binary system to be freely falling, hence to leading order its center of mass moves on a geodesic of the background space-time. See \cite{Maeda:2023tao} for a similar analysis.
\lbreak

In general, we can describe the metric near a geodesic of a background space-time using the Thorne-Hartle version of the Fermi-normal coordinates \cite{PhysRevD.31.1815}
\begin{equation}
\label{FNmetric}
\begin{split}
    &g_{00} = - 1 - \mathcal{E}_{ij} x^i x^j + \mathcal{O} ((x/\mathcal{R})^3)~~,
\\   &g_{0i} = - \frac{2}{3} \epsilon_{ijk} \mathcal{B}^j{}_l x^k x^l + \mathcal{O} ((x/\mathcal{R})^3)~~,
\\    &g_{ij} = \delta_{ij} ( 1 -  \CE_{kl} x^k x^l ) + \CO ((x/\mathcal{R})^3)~~,
\end{split}
\end{equation}
where $i,j=1,2,3$, the square of the geodesic distance to the geodesic is $x^2 = x^i x^i$ and $\mathcal{R}$ is the curvature length scale. 
To use the tidal force description of the background we need that the curvature length scale $\mathcal{R}$ of the background is significantly larger than the size $x$ of the binary system, hence $x \ll \mathcal{R}$~\cite{Poisson:2009qj}. This is known as the  ``small-tide approximation''.
\lbreak

That we keep terms up to order $x^2/\mathcal{R}^2$ in \eqref{FNmetric} encaptures the quadrupole approximation of the tidal forces, for which the electric and magnetic tidal moments $\CE_{ij}$ and $\CB_{ij}$ are related to the Riemann curvature tensor as
\begin{equation}
    \CE_{ij} =  R_{0i0j}|_{x=0}~~~,~~~
    \CB_{ij} = \frac{1}{2} \epsilon_{pq (i} R^{pq}|_{j)0}|_{x=0}~~,
\end{equation}
with $i,j,p,q=1,2,3$.
Note that these quantities depend in general on the proper time $\tau$ of the geodesic. We assume that the background space-time has $R_{\mu\nu}=0$ which is indeed true for the Kerr space-time.
Below in Section \ref{sec:binary_kerr} we will restrict ourselves to tidal moments $\CE_{ij}$ and $\CB_{ij}$ for geodesics around a Kerr black hole, as computed in \cite{Marck:1983,Camilloni:2023rra}, but for the moment we keep our considerations general.
\lbreak

We now model the two BHs in the binary system as two particles moving in the background metric \eqref{FNmetric} with the center of mass of the binary located at the geodesic. 
The Lagrangian for particle 1 is
\begin{equation}
\label{Lagrfct}
L_{(1)} = - m_1 c^2 \sqrt{-G_{00} - 2G_{0i} \frac{v_{(1)}^i}{c}- G_{ij}\frac{v_{(1)}^i v_{(1)}^j}{c^2}}~.
\end{equation}
Here $v_{(1)}^i$ is the velocity of particle 1.
The metric $G_{\mu\nu}$ is evaluated at the position $\vec{x}_1$ of particle 1 and it consists
of the background metric $g_{\mu\nu}$ with tidal forces \eqref{FNmetric} plus the gravitational potentials generated by particle 2. For the latter, we use the PN expansion of Sections 5.1.4 and 5.1.5 in the book \cite{Maggiore:2007ulw}. Explicitly, we have up to 1PN
\begin{equation}
\begin{split}
G_{00} &= g_{00} - 2 \phi_{(2)} - 2 (\phi_{(2)}^2 + \psi_{(2)} ) + \cdots
\\
G_{0i} &=g_{0i} + \zeta_i^{(2)}+ \cdots
\\
G_{ij}&=g_{ij} -2 \delta_{ij} \phi_{(2)}+ \cdots
\end{split}
\end{equation}
where the standard PN gravitational potentials $\phi_{(2)}$, $\psi_{(2)}$ and $\zeta_i^{(2)}$ arise solely from the gravitational interaction with particle 2, see \cite{Maggiore:2007ulw} for details. Here we are neglecting possible terms that correspond to mixed couplings between the tidal forces and the PN expansion. Note that the first appearance would be terms of the type $\phi_{(2)}$ times the tidal moments.
\lbreak

One can now perform a PN expansion of the Lagrangian \eqref{Lagrfct} for particle 1, with the restriction that the tidal moments provide a small perturbation that only should be included up to the first order. 
Schematically, this gives
\begin{equation}
\label{LagrParticle1}
\begin{split}
    L_{(1)} =& L_{(1)}|_{\mathcal{E}=\CB=0} - \frac{1}{2} m_{1} c^2 x_{(1)}^i x_{(1)}^j \mathcal{E}_{ij} 
    \\ 
    &- \frac{2c}{3}m_{1}   v_{(1)}^i x_{(1)}^k x_{(1)}^l \epsilon_{ijk} \mathcal{B}^j{}_l + \cdots
\end{split}
\end{equation}
Here the first term $L_{(1)}|_{\mathcal{E}=\mathcal{B}=0}$ is the Lagrangian that arises solely from the gravitational interaction with particle 2, as well as the kinetic energy of particle 1.
\lbreak

The second term in \eqref{LagrParticle1}, coupling to $\CE_{ij}$, is the leading coupling to the quadrupole tidal moments of the background space-time. 
\lbreak

The third term in \eqref{LagrParticle1}, coupling to $\CB_{ij}$, is subleading as it is suppressed by $v_{(1)}^i/c$. Note that  the coupling between the gravitational potentials and the tidal moments would be of order $v_{(1)}^2/c^2$ or higher, thus not affecting this term. 
\lbreak

Restricting for a moment ourselves to the Newtonian limit (with respect to particles 1 and 2) of all of the terms, and adding the analogous Lagrangian for particle 2, we get the Lagrangian for the Newtonian limit of the binary system
\begin{align}
\nonumber
L_{\rm Newton} =& - (m_1+m_2) c^2 + \frac{1}{2} m_1 v_{(1)}^2 + \frac{1}{2} m_2 v_{(2)}^2 
\\\label{L_newton1}
&+ \frac{G m_1 m_2}{r}- \frac{c^2}{2} \Big[ m_{1}  x_{(1)}^i x_{(1)}^j+ m_{2}  x_{(2)}^i x_{(2)}^j \Big] \mathcal{E}_{ij}.
\end{align}
Introducing now the center of mass  quantities
\begin{equation}
\label{eq: comframe}
\begin{split}
&M = m_1+m_2 \spa \mu = \frac{m_1m_2}{M}~~,
    \\
&\vec{X} = \frac{m_1 \vec{x}_{(1)} + m_2 \vec{x}_{(2)}}{M} \spa \vec{x} = \vec{x}_{(2)}-\vec{x}_{(1)} \spa r = |\vec{x}|~~,
    \\
&\vec{V} = \frac{m_1 \vec{v}_{(1)} + m_2 \vec{v}_{(2)}}{M} \spa \vec{v} = \vec{v}_{(2)}-\vec{v}_{(1)} ~~,
\end{split}
\end{equation}
the Lagrangian \eqref{L_newton1} becomes
\begin{align}
\nonumber
 L_{\rm Newton} =& - M c^2 + \frac{1}{2} M V^2 + \frac{1}{2} \mu v^2  
 \\\label{L_newton2}
    &+ \frac{G M \mu}{r}- \frac{c^2}{2} \Big[ M  X^i X^j+ \mu   x^i x^j \Big] \mathcal{E}_{ij}~~.
\end{align}
Since the center of mass motion of the binary is decoupled from the relative motion of the binary system we can consistently set 
\begin{equation}
\vec{X}=0 \spa \vec{V} = 0~~.
\end{equation}
This means we put the center of mass of the binary system at the geodesic, which is what one physically would imagine as well (for large distance scales the binary system should be seen as one particle of mass $M$ placed at the geodesic).
Using moreover the general formula (\cite{Poisson:2009qj})
\begin{equation}\label{Poisson}
x^i x^j   \mathcal{E}_{ij} = r^2 \mathcal{E}^q ~~,
\end{equation}
where $\mathcal{E}^q$ is the quadrupole tidal potential,
we get the binary system Lagrangian
\begin{equation}
\label{L_newton3}
 L_{\rm Newton} =   \frac{1}{2} \mu v^2  
+ \frac{G M \mu}{r}- \frac{c^2}{2} \mu \,  r^2 \mathcal{E}^q~~.
\end{equation}
One can now easily Legendre transform this to the Hamiltonian
\begin{equation}
\label{H_newton}
 H_{\rm Newton} =   \frac{1}{2\mu} p^2  
- \frac{G M \mu}{r}+ \frac{c^2}{2} \mu \,  r^2 \mathcal{E}^q~~,
\end{equation}
with $p^i=\mu v^i$.
This Hamiltonian describes the general situation of a binary system of particles moving on a geodesic of a background space-time subject to quadrupole tidal forces from the curvature of the background space-time. This is valid in a local inertial system as set by the Fermi-normal coordinates \eqref{FNmetric}. 
Below in Section \ref{sec:gyroscope} we shall explore the same dynamics in a non-inertial system as well.

\section{Binary system in the background of large Kerr black hole}
\label{sec:binary_kerr}

We consider now a system of three BHs of masses $m_1$, $m_2$ and $m_3$. The two BHs with masses $m_1$ and $m_2$ are in a bound motion and constitute the BBH system. 
We will often refer to it as the inner binary, and their motion as the inner orbit. Their masses are taken to be much smaller than the mass of the third BH, namely $m_1,m_2 \ll m_3$, thus identifying $m_3$ as the SMBH.  
\lbreak

We will assume two independent separations of scales, in order to gain analytical control of the dynamics. The first one is that the two BHs in the inner binary are in the particle/PN regime for which their Schwarzschild radii are much smaller than their separation $r$: 
\begin{equation}
\label{SEPbinary}
r \gg \frac{2Gm_1}{c^2}~,~~ \frac{2Gm_2}{c^2}~~.
\end{equation}
In particular,  their relative velocity is much smaller than the speed of light.
\lbreak

Secondly, we assume the small-tide approximation, {\sl i.e.}~that the size of the binary system, $r$, is very small compared to the curvature length scale $\mathcal{R}$ of the SMBH. 
This enables us to treat the influence of the SMBH on the BBH via a tidal force approximation, even when the BBH is close to the SMBH.
Under these assumptions one has that the motion of the BBH, to leading order, is a geodesic of the Kerr metric describing the SMBH. This we will refer to as the outer orbit.  
On the geodesic the small-tide approximation $r \ll \mathcal{R}$ can be written explicitly as
\begin{equation}
\label{SEPtidal}
r \ll \hat{r} \sqrt{\frac{c^2 \hat{r}}{G m_3}}~~,
\end{equation}
with $\hat r$ the radial coordinate of the Kerr metric that we introduce below. 
\lbreak

We will restrict ourselves in this paper to the leading quadrupole effect, arising from the Riemann curvature tensor of the Kerr metric evaluated on the geodesic. Below we shall include PN effects in the binary dynamics, in the form of the periastron precession and the GW radiation-reaction, thus we assume that the octupole tidal forces are smaller than these PN effects.

\subsection{Kerr black hole background} 

In our setup we consider the SMBH to be a Kerr black hole, whose line element, represented in Boyer-Lindquist coordinates $\hat x^\mu=(\hat t,\hat r,\hat \theta,\hat\phi)$, reads
\begin{equation} 
    \label{KerrBL}
\begin{split}
    d\hat s^2=&-\Big(1-\frac{2Gm_3\hat r}{c^2\Sigma}\Big)c^2d\hat t^2-\frac{4Gm_3\hat r}{c^2\Sigma}s_3\sin^2 \hat\theta ~c\, d\hat t\, d\hat \phi\\
    &+\frac{\mathcal{A}}{\Sigma}\sin^2 \hat \theta ~d\hat \phi^2
    +\frac{\Sigma}{\Delta}d\hat r^2+\Sigma d\hat \theta^2~~.
\end{split}
\end{equation}
In the metric above, 
$m_3$ is the black hole mass, $s_3=J_3/(c\, m_3)$ is the specific angular momentum and we defined
\begin{equation}
\label{sigma}
\begin{split}
    \Sigma = \hat r^2+&s_3^2 \cos^2 \hat \theta~, ~~ \Delta=\hat r^2- \frac{2Gm_3}{c^2} \hat r+s_3^2 ~~,
    \\
    \mathcal{A} &= (\hat r^2+s_3^2)^2-\Delta s_3^2\sin^2 \hat \theta~~.
\end{split}
\end{equation}
In the following it is convenient to introduce the dimensionless spin parameter 
$\chi$
\begin{equation}
    \label{alpha}
    \chi=\frac{s_3 c^2}{G m_3}~~.
\end{equation}
The location of the event horizon is given by the major root of the equation $\Delta=0$ and the cosmic censorship conjecture requires $0\leq s_3\leq (G m_3/c^2)$, $i. e.$ $0\leq\chi\leq 1$.\\
\lbreak

A generic geodesic $\hat{x}^\mu(\tau)$ in the Kerr space-time is parametrized by three constants of motion, respectively representing the energy $\hat E$, the angular momentum $\hat L$ and the Carter constant $K$ per unit of rest energy \cite{PhysRev.174.1559}.
\lbreak

Upon restricting to the circular ($\hat{r}$ constant) and equatorial ($\hat \theta=\pi/2$) case, the tangent vector $u^\mu$, representing the four-velocity of a particle moving along the geodesic can be written as 
\begin{equation}
    u^\mu\equiv\frac{d\hat x^\mu}{d\tau}=u^t(\delta^\mu_t+\Omega~\delta^\mu_\phi)~~,
\end{equation}
where the redshift factor $u^t$ and the coordinate angular velocity $\Omega$ are defined as
\begin{equation}
\label{eq: redshift}
\begin{split}
    u^t\equiv\frac{d\hat t}{d\tau}&=\frac{1}{\sqrt{-(g_{tt}+2\Omega g_{t\phi}+\Omega^2 g_{\phi\phi})}}~~,
    \\
    \Omega\equiv\frac{d\hat\phi}{d\hat t}&=\frac{\sigma\left({Gm_3}\right)^{1/2}}{\hat r^{3/2}+\sigma s_3 \left(\frac{Gm_3}{c^2}\right)^{1/2}}~~,
    \end{split}
\end{equation}
and where we use $\sigma=\pm1$ to respectively distinguish orbits that are co-rotating and counter-rotating relative to the angular momentum of the Kerr black hole.~\footnote{In the Schwarzschild limit $\chi=0$ then $\sigma$ distinguishes anti-clockwise and clockwise orbits, respectively.}
For later convenience, we also report the orbital angular velocity defined with respect to the proper time, given by $\Omega_{\hat\phi}=u^t\Omega$, namely
\begin{equation}
\label{eq: omega_0}
    \Omega_{\hat\phi}\equiv\frac{d\hat \phi}{d\tau}=\frac{\sigma\left(G m_3\right)^{1/2}}{\hat r^{1/2}\sqrt{\hat r^2+2\sigma s_3 \left(\frac{Gm_3}{c^2}\hat r\right)^{1/2}-3\frac{Gm_3}{c^2} \hat{r}}}~~. 
\end{equation}
In this case it is possible to obtain explicit expressions for the constants of motion in terms of the orbital radius $\hat r$, the black hole mass $m_3$, and the specific spin value $s_3$. More specifically, one has the energy, orbital angular momentum, and Carter constant per unit of rest energy
\begin{align}
\label{eq: ELK_Kerr}
\begin{split}
    \hat E &=\frac{\hat{r}^{3/2}-\frac{2Gm_3}{c^2} \hat{r}^{1/2} + \sigma s_3  \left(\frac{Gm_3}{c^2}\right)^{1/2}}{\hat{r}^{3/4} \sqrt{\hat{r}^{3/2}-3\frac{Gm_3}{c^2} \hat{r}^{1/2} + 2\sigma s_3   \left(\frac{Gm_3}{c^2}\right)^{1/2} }}~~,
    \\
    \hat L &=\frac{\sigma  \left(\frac{Gm_3}{c^2}\right)^{1/2} \left(\hat{r}^2+  s_3 ^2  -2 \sigma s_3  ~\left(\frac{Gm_3}{c^2}\hat{r}\right)^{1/2}  \right)}{\hat{r}^{3/4} \sqrt{\hat{r}^{3/2}-3\frac{Gm_3}{c^2} \hat{r}^{1/2} + 2\sigma s_3   \left(\frac{Gm_3}{c^2}\right)^{1/2} } }~~,
    \\
    K &= \left(s_3   \hat E-\hat L \right)^2=\left(\hat r-\sigma s_3 \sqrt{\frac{c^2 \hat r}{G m_3}}\right)^2\frac{ \hat r^2\Omega_{\hat{\phi}}^2}{c^2}~~.
\end{split}
\end{align}

For later use, we introduce here the ISCO, which identifies the innermost stable circular orbit. In the equatorial plane of a Kerr black hole there exist two ISCOs, one of which is co-rotating ($\sigma=+1$) and one which is counter-rotating ($\sigma=-1$) with the black hole. The value of the radial coordinate at which the ISCOs are located in the equatorial plane of the Kerr space-time is given by \cite{1972ApJ...178..347B}
\begin{equation}
    \label{eq:ISCOKERR}
\hat{r}^\sigma_{\rm ISCO}= \frac{G m_3}{c^2}\left[3+Z_2-\sigma \sqrt{(3-Z_1)(3+Z_1+2Z_2)}\right]~,
\end{equation}    
where
\begin{equation}
\begin{split}
    Z_1 &=1+\left(1-\chi^2\right)^{\frac{1}{3}}\left[\left(1+\chi\right)^{\frac{1}{3}}+\left(1-\chi\right)^{\frac{1}{3}}\right]~,
    \\
    Z_2&=\sqrt{Z_1^2+3\chi^2}~.
\end{split}
\end{equation}
One can see immediately that for $\chi=0$ eq.~\eqref{eq:ISCOKERR} reduces to $\hat r_{\rm ISCO}=6 G m_3/c^2$, i.e. the usual ISCO position for a Schwarzschild black hole. 

\subsection{Binary system on a Kerr geodesic}
\label{sec:marck_tetrad}

As explained above, we can regard the black holes of the BBH system as particles in a Newtonian limit since we assume the condition \eqref{SEPbinary} holds. The BBH system moves along a circular equatorial orbit in the background of a Kerr black hole describing the SMBH. As seen in Section \ref{sec:binary_general}, for a sufficiently small binary system \eqref{SEPtidal} we can approximate the influence of the SMBH through quadrupole tidal forces acting on the BBH, while the center of mass of the BBH moves approximately on a geodesic of the Kerr metric. This means we can employ the results of Section \ref{sec:binary_general}.
\lbreak

For the Kerr metric \eqref{KerrBL}, we can use the {\it{Marck's tetrad}} \cite{Marck:1983} to describe
the local Fermi-normal coordinates \eqref{FNmetric} of Section \ref{sec:binary_general}. 
Marck's tetrad is given by the four vectors $\lambda_A^\mu$ which provide an orthonormal basis for the vector space at each point of the Kerr-geodesic since $\lambda_A^\mu \lambda_B^\nu g_{\mu\nu} = \eta_{AB}$ and $\lambda_A^\mu \lambda_B^\nu \eta^{AB} = g_{\mu\nu}$, where in particular $\lambda_0^\mu = u^\mu$ is the four-velocity. 
Note that $A=0,1,2,3$ are the flat tetrad indices.
One can equivalently represent  Marck's tetrad as the orthonormal one-forms $\lambda^A_\mu = \eta^{AB} g_{\mu\nu} \lambda_B^\nu$.
\lbreak

One can now employ the standard  map of the Fermi-Normal coordinates between vectors on the Kerr geodesic and events in the neighborhood of the geodesic. For Marck's tetrad, we have in addition to the time coordinate $x^0=\tau$, which is the proper time on the geodesic, also the spatial coordinates $x^i$ parametrizing an orthogonal vector $\sum_{i=1}^3 x^i \lambda^\mu_i$ at the geodesic. In this way the coordinates $x^A$ describe a neighborhood of the geodesic. We will call this coordinate system the {\sl Marck's frame of reference}.
\lbreak

Since Marck's tetrad is parallel-transported along the geodesic, it provides an inertial frame, meaning it is characterized by a vanishing acceleration and vanishing angular velocity of rotation of spatial basis vectors \cite{Misner1973} \footnote{Both formulas can be derived starting from the general transport law for an observer's tetrad, $D \lambda^\mu_{a}/d\tau=-{\Omega^\mu}_\nu \lambda^\nu_a$, with the quantity $\Omega_{\mu\nu}=a_\mu u_\nu-u_\mu a _\nu+u^\alpha \omega^\beta \epsilon_{\alpha \beta \mu \nu}$ \cite{Misner1973}.} 
\begin{equation}
\label{eq: Marck_aw}
    a^i\equiv\lambda^i_\mu\frac{D\lambda^\mu_{0}}{d\tau}=0~~~,
    ~~~
    {\omega}^i\equiv-\frac{1}{2}{\epsilon^{ij}}_{k}{{\lambda}}^\mu_j\frac{D{{\lambda}}_\mu^k}{d\tau}=0~~.
\end{equation}
For circular geodesics in the equatorial plane ($\hat \theta=\pi/2$), Marck's tetrad can be conveniently written in the one-forms basis as 
\begin{equation}
\label{eq: MarckT}
    \begin{split}
        \lambda_\mu^{0}&=\left(\hat E,0,0,\hat L\right)~~,
        \\
        {\lambda}_\mu^{1}&=\cos\Psi ~\tilde \lambda^{1}_\mu-\sin\Psi~ \tilde \lambda^{2}_\mu~~,
   \\
        {\lambda}_\mu^{2}&=\sin\Psi ~\tilde \lambda^{1}_\mu+\cos\Psi~ \tilde \lambda^{2}_\mu~~,
\\
        \lambda_\mu^{3}&=\left(0,0,-\hat r,0\right)~~,
\end{split}
\end{equation}
with
\begin{equation}
    \begin{split}
        \tilde{\lambda}_\mu^{1}&=\left(0,\sqrt{\frac{\hat r^2}{K+\hat r^2}}\frac{(\sqrt{K}s_3+\hat E \hat r^2)}{\Delta},0,0\right)~~,
        \\
        \tilde{\lambda}_\mu^{2}&=\left(\frac{(s_3-\hat E\sqrt{K})}{\sqrt{K+\hat r^2}},0,0,\frac{\hat E \sqrt{K}s_3-s_3^2-\hat r^2-K}{\sqrt{K+\hat r^2}}\right)~~,
    \end{split}
\end{equation}
where $\hat r$ is the constant radius of the circular equatorial geodesic and the angle $\Psi$ is introduced to ensure that the tetrad is parallel transported along the geodesic \cite{Marck:1983,Camilloni:2023rra}, as shown by Eqs.~\eqref{eq: Marck_aw}. The explicit expression for $\Psi$ in terms of the geodesic's proper time $\tau$ is given by $\Psi=\Omega_\Psi \tau$, with
\begin{equation}
    \label{eq: omega_r}
    \Omega_\Psi\equiv\sigma \sqrt{\frac{G m_3}{\hat r^3}}~~.
\end{equation}

To use the results for the Lagrangian and Hamiltonian of Section \ref{sec:binary_general}, given in Eqs.~\eqref{L_newton3} and \eqref{H_newton}, respectively, we record that the electric tidal moments in the equatorial plane are \cite{Marck:1983,Camilloni:2023rra}
\begin{equation}
\label{eq: C_ij}
\begin{split}
    \mathcal{E}_{11}&=\left[1-3\left(1+\frac{K}{\hat r^2}\right)\cos^2\Psi\right]\frac{G m_3}{c^2\hat r^3}~~,
    \\
    \mathcal{E}_{22}&=\left[1-3\left(1+\frac{K}{\hat r^2}\right)\sin^2\Psi\right]\frac{G m_3}{c^2\hat r^3}~~,
    \\
    \mathcal{E}_{33}&=\left(1+3\frac{K}{\hat r^2}\right)\frac{G m_3}{c^2\hat r^3}~~,
    \\
   \mathcal{E}_{12}&=-3\left(1+\frac{K}{\hat r^2}\right)\frac{G m_3}{c^2\hat r^3}\cos\Psi\sin\Psi~~.
\end{split}
\end{equation}
Using these with Eq.~\eqref{Poisson}, one gets an explicit expression for the scalar quadrupole electric tidal moment induced by the Kerr black hole, as measured by an observer using Fermi-normal coordinates, namely
\begin{equation}
\label{eq: tidal_pot}
\begin{split}
    r^2\mathcal{E}^q
    =\frac{G m_3}{c^2 \hat r^3}&\left[r^2+3(x^3)^2\frac{K}{\hat{r}^2}\right.
    \\
    &\left.-3\left(1+\frac{K}{\hat r^2}\right)(x^1\cos\Psi+x^2\sin\Psi )^2\right]~~.
\end{split}
\end{equation}
With this, we can describe the dynamics of the binary system in the approximations \eqref{SEPbinary} and \eqref{SEPtidal} via the Lagrangian \eqref{L_newton3} and Hamiltonian \eqref{H_newton} of the BBH system.

\section{Gyroscope precession and the distant-star frame}
\label{sec:gyroscope}

In this section, we consider the gyroscope precession of Marck's parallel transported frame of reference by introducing a non-inertial frame of reference that we dub the distant-star frame. %
\footnote{The gyroscope precession was also considered for a BBH system in orbit around a Schwarzschild black hole in \cite{Maeda:2023tao}.}
\lbreak

The BBH has an orbital angular momentum that undergoes a gyroscope precession in its motion along an equatorial circular geodesic in Kerr space-time around
the SMBH. This is induced by the curvature of the background, like in the case of the Earth-Moon binary system in orbit around the Sun~\cite{Rindler}. 
In the Schwarzschild space-time this gyroscope precession is known as the \textit{Fokker-de Sitter precession}~\cite{1917MNRAS..78....3D}, whereas in the equatorial plane of a  Kerr black hole, it takes the name of \textit{Schiff's precession}~\cite{PhysRevLett.4.215}.
\lbreak

The origin of this precession is the difference between the local and global points of view for our binary system moving on a geodesic. Marck's frame represents the local view point, where we have an approximate inertial system close to the center of mass of the binary that moves on the geodesic. However, there is also a global point of view, in which the global properties of the Kerr space-time are taken into account. 
This is clear in our case of an equatorial circular geodesic. For this motion, the only spatial coordinate in the BL coordinates of Kerr that changes is the angle $\hat{\phi}$ as
\begin{equation}
\label{hatphi}
\hat{\phi}=\Omega_{\hat{\phi}} \tau~~,
\end{equation}
where $\Omega_{\hat{\phi}}$ is given in \eqref{eq: omega_0}.
A period of motion is obviously when $\hat{\phi}$ changes with $2\pi$. However, Marck's frame is not the same after one period, since the $\Psi$ angle has changed with $\Delta \Psi = 2\pi (\Omega_{\Psi}/\Omega_{\hat{\phi}} -1)$, which gives the gyroscope precession.
\lbreak

Since the precession is not observable in Marck's frame of reference by itself, it is useful to define a non-inertial frame of reference in which the gyroscope precession is manifest. Such a frame, here called the {\sl distant-star} frame of reference, is constructed in \cite{Bini:2016iym},
simply by rotation of Marck's frame with angle $\hat{\phi}-\Psi$ such that the distant-star frame is periodic under rotations with respect to the $\hat{\phi}$ angle.
Specifically, seeing it as a tetrad $e^i_\mu$, it is defined by the following rotation of Marck's tetrad $e^{i}_\mu={R^i{}_j} \lambda^{j}_\mu$ with
\begin{equation}
    \label{tetrad_rot}{R^i{}_j}=\begin{pmatrix}
    \cos (\Omega_{\rm g} \tau) & -\sin (\Omega_{\rm g} \tau) & 0
    \\
    \sin(\Omega_{\rm g} \tau) &\cos(\Omega_{\rm g} \tau) &0
    \\
    0 &0 &1
    \end{pmatrix}~~.
\end{equation}
where we introduced the gyroscope angular velocity $\Omega_{\rm g}$ as
\begin{equation}
\label{eq: Omegag}
    \Omega_{\rm g}=\Omega_{\hat\phi}-\Omega_\Psi=\Omega_\Psi\left(\frac{1}{\sqrt{1+2 \frac{s_3}{c}\Omega_\Psi-3\frac{{\hat r^2}}{c^2}\Omega_\Psi^2}} -1\right)~~,
\end{equation}
such that $\hat{\phi}-\Psi = \Omega_{\rm g} \tau$. It is easy to check that $\Omega_g=0$ at $\hat{r}=4/9~(G m_3/c^2) \chi^2 $, and that this location never corresponds to a stable orbital radius, since $\hat{r}_\star<\hat r^\sigma_{\rm ISCO}$ for $0\leq\chi\leq1$. For completeness, we report that at the ISCO one has $\Omega_g=(c/\hat r_{\rm ISCO}) (\sqrt{2}-1)/\sqrt{6}$ in the Schwarzschild case $\chi=0$, whereas in the extreme Kerr case, $\chi=1$, the gyroscope precession diverges as $\Omega_g\approx2 c/\sqrt{3} (\hat r- \hat r^+_{\rm ISCO})^{-1}$ for co-rotating orbits, and $\Omega_g=(c/\hat r^-_{\rm ISCO})(1/3-\sqrt{3}/4)$ for the counter-rotating ones. 
As we shall see below, the distant-star frame provides a local coordinate system close to the circular equatorial geodesic in which one can directly observe the precession as a fictitious force in the Lagrangian description. 
\lbreak

In general, a local observation of a precession angle is not possible, since one cannot compare angles between two events in  space-time in a path-independent manner. However, the construction of the distant-star tetrad is based on the global structure of the Kerr space-time, being stationary and axisymmetric, which gives a natural definition of angular and radial directions in the equatorial plane through Carter's tetrad \cite{Bardeen:1972fi}. Thus, in this sense, one can meaningfully claim the distant-star frame is fixed with respect to the asymptotic definition of the rotating angle, justifying its name as an angle with respect to distant stars. {\sl I.e.}~the distant-star frame provides a Cartesian frame that keeps a fixed orientation with respect to distant stars \cite{Bini:2016iym}. 
Hence, the non-inertial distant-star frame of reference provides a global point of view, contrary to the local inertial Marck's frame of reference.
\lbreak

One has that the distant-star tetrad $e^{i}_\mu$ is characterized by a vanishing acceleration, being anchored to the geodesic, but a non-vanishing angular velocity of rotation relative to the Marck's tetrad \cite{Misner1973}
\begin{equation}
\label{eq: distant_aw}
    a^i\equiv e^i_\mu\frac{D{e}_0^\mu}{d\tau}=0~~~,
    ~~~ \omega^i\equiv-\frac{1}{2}{\epsilon^{ij}}_k e^\mu_j\frac{De_\mu^k}{d\tau}=\Omega_g \delta_3^i~~,
\end{equation}
where we used the vector tetrad basis $e^\mu_i = \lambda^\mu_j{(R^T)^j}_i$ and $e^\mu_0=\lambda^\mu_0=u^\mu$.
\lbreak

The spatial coordinates $\boldsymbol{r}^i$ associated with the distant-star tetrad are given by
\footnote{This follows from the fact that for any vector $V^\mu$  its spatial components are  $\boldsymbol{r}^i = V^\mu e^i_\mu = V^\mu R^i{}_j \lambda^j_\mu$ and $x^i = V^\mu \lambda^i_\mu$.}
\begin{equation}\label{Binirotation}
\boldsymbol{r}^i = R^i{}_j x^j~~.
\end{equation}
In the following, we shall use the Cartesian vector notation
\begin{equation}
\label{pos_vec}
\boldsymbol{r} = \boldsymbol{r}^1 \boldsymbol{\hat{x}}+ \boldsymbol{r}^2 \boldsymbol{\hat{y}}+ \boldsymbol{r}^3\boldsymbol{\hat{z}}~~,
\end{equation}
where we defined the unit vectors
\begin{equation}
\boldsymbol{\hat{x}} = \begin{pmatrix}    1    \\    0    \\    0  \end{pmatrix}~~,~~
\\
\boldsymbol{\hat{y}} = \begin{pmatrix}    0    \\    1    \\    0  \end{pmatrix}~~,~~
\\
\boldsymbol{\hat{z}} = \begin{pmatrix}    0    \\    0    \\    1  \end{pmatrix}
~~.
\end{equation}
The consequence of going to this non-inertial frame for the binary system is the introduction of fictitious forces. Indeed, the local Lagrangian now becomes
\begin{equation}
\begin{split}
    \mathcal{L}=&\frac{\mu}{2}(\boldsymbol{v}-\boldsymbol{\Omega_{\rm g}}\times \boldsymbol{r})^2+\frac{G M\mu}{r}-\frac{c^2\mu}{2}r^2\mathcal{E}^q~~,
\end{split}
\end{equation}
where we introduced the Cartesian vectors $\boldsymbol{v}= d\boldsymbol{ r}/d\tau$ and $\boldsymbol{\Omega_{\rm g}} = \Omega_{\rm g} \boldsymbol{\hat{z}}$ in agreement with Eq.~\eqref{eq: distant_aw}.
Now Eq.~\eqref{Poisson} reads
\begin{equation}
\label{eqnoninertial}
\begin{split}
    r^2\mathcal{E}^q
    =&\frac{G m_3}{c^2 \hat r^3}\left[r^2+3(\boldsymbol{r}^3)^2\frac{K}{\hat{r}^2}+\right.
    \\
    &\left.-3\left(1+\frac{K}{\hat r^2}\right)(\boldsymbol{r}^1\cos\hat{\phi}+\boldsymbol{r}^2\sin\hat{\phi})^2\right]~~.
\end{split}
\end{equation}

To find the Hamiltonian we define the canonical momentum as
\begin{equation}
    \boldsymbol{\pi}=\frac{\partial\mathcal{L}}{\partial \boldsymbol v}=\mu\left(\boldsymbol{v}- \boldsymbol{\Omega}_{\rm g}\times\boldsymbol{r}\right)~~,
\end{equation}
and the canonical angular momentum as
\begin{equation}
    \boldsymbol{L}_{\rm in}=\boldsymbol{r}\times \frac{\partial\mathcal{L}}{\partial \boldsymbol v}=\boldsymbol{r}\times  \boldsymbol{\pi}~~,
\end{equation}
where we adopt the subscript ``\rm{in}" for later convenience  to distinguish the angular momentum of the inner BBH system and the angular momentum associated with the outer orbit.
The Hamiltonian, thus, reads 
\begin{equation}\label{HBini}
\begin{split}
    \mathcal{H}=&\frac{\boldsymbol{\pi}^2}{2\mu}-\frac{G\mu M}{r}+\boldsymbol{\Omega}_{\rm g}\cdot\boldsymbol{L}_{\rm in} +\frac{\mu c^2}{2}r^2\mathcal{E}^q~~,
\end{split}
\end{equation}
with $\mathcal{E}^q$ given in Eq.\eqref{eqnoninertial}. The extra term in \eqref{HBini} with respect to \eqref{H_newton} is responsible for the gyroscope precession of $\boldsymbol{L}_{\rm in}$.

\section{Euler angles and action-angle variables}
\label{sec: euler_angles_and_action_angle_variables}

In Sections \ref{sec:angles} and \ref{sec:actions}, we review certain standard definitions of  angular coordinates and momenta that are highly useful to describe the inner binary, and to derive the secular Hamiltonian in Section \ref{sec: <H>}. All of these definitions are with respect to the distant-star frame of reference introduced in Section \ref{sec:gyroscope}.
Subsequently, in Section \ref{sec:marck_variables} we introduce for later use the action-angle variables in Marck's frame of reference as well, and find the canonical transformation between the distant-star and Marck's frame Hamiltonians.

\subsection{Euler angles}
\label{sec:angles}

While the center of mass for the inner Newtonian BBH system moves along the circular equatorial geodesic in the Kerr space-time, the  orientation of its inner orbital plane can vary with respect to the outer orbital plane of the Kerr SMBH.  
\lbreak

The vector \eqref{pos_vec} describing the relative position of a body in a Newtonian elliptic orbit can be represented as
\begin{equation}
\label{ruv}
    \boldsymbol{r}=r (\cos\psi~ \boldsymbol{\hat u}+\sin\psi ~\boldsymbol{\hat v})~~,
\end{equation}
where
\begin{equation}
    \label{eq: r_ea}
    r=\frac{a (1-e^2)}{1+e \cos\psi}~~,
\end{equation}
where $a$ and $e$ are respectively  the \textit{semi-major axis} and the \textit{eccentricity} of the orbit, whereas $\psi$ is the angular coordinate that keeps track of the body motion along the orbit, namely the \textit{true anomaly}.
\lbreak

The two directions $\boldsymbol{\hat u}$ and $\boldsymbol{\hat v}$ have a precise geometrical meaning, with $\boldsymbol{\hat u}$ ($\psi=0$) identifying the \textit{periapsis direction}, and $\boldsymbol{\hat v}$ ($\psi=\pi/2$) lying along the \textit{direction of the ascending nodes}. The space spanned by these two vectors specifies the inner orbital plane.
Including the direction of the angular momentum for the inner binary  $\boldsymbol{\hat L}_{\rm in}=\boldsymbol{\hat u}\times \boldsymbol{\hat v}$, one  obtains a triad of orthonormal vectors.

Since the inner binary is assumed to be in a Newtonian regime, it is possible to unambiguously introduce the \textit{eccentric anomaly} $\zeta$ and the \textit{mean anomaly} $\beta$, defined by means of 
\begin{equation}
\label{beta_psi}
    \cos\psi=\frac{\cos\zeta-e}{1-e \cos\zeta}~~~,~~~\beta=\zeta-e\sin \zeta~~.
\end{equation}
The first relation defines $\zeta$ in terms of the true anomaly $\psi$, whereas the second is the Kepler equation. Kepler's equation is a transcendental equation and no closed-form solution is known that allows the expression of the eccentric anomaly $\zeta$ in terms of the mean anomaly $\beta$.
\lbreak

The mean anomaly $\beta$ is particularly useful as it represents the angle that a fictitious body moving in a circular orbit would span if it had the same orbital frequency as the actual body moving along the elliptic orbit. In other words its motion is uniform in time, $\beta=\sqrt{G M /a^3}\tau $.
\lbreak

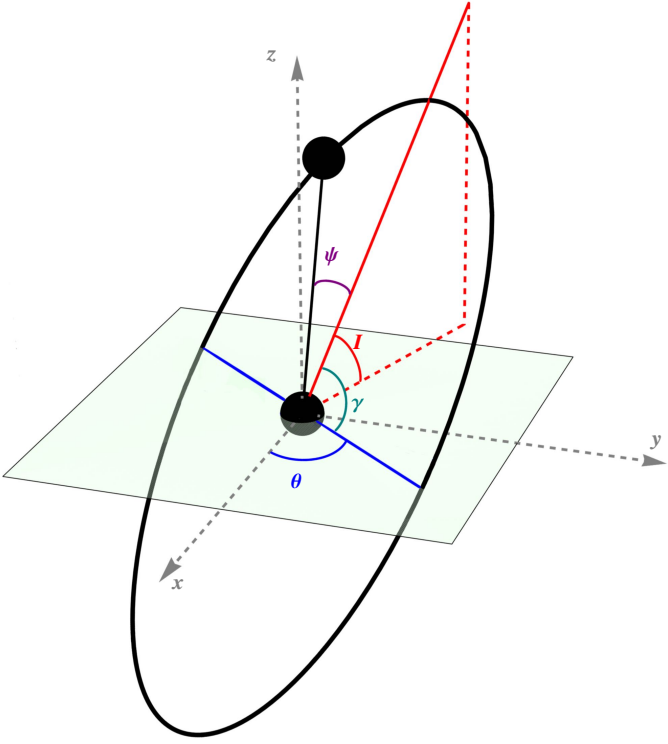
\begin{figure}
\centering

\includegraphics[width=0.50\textwidth]{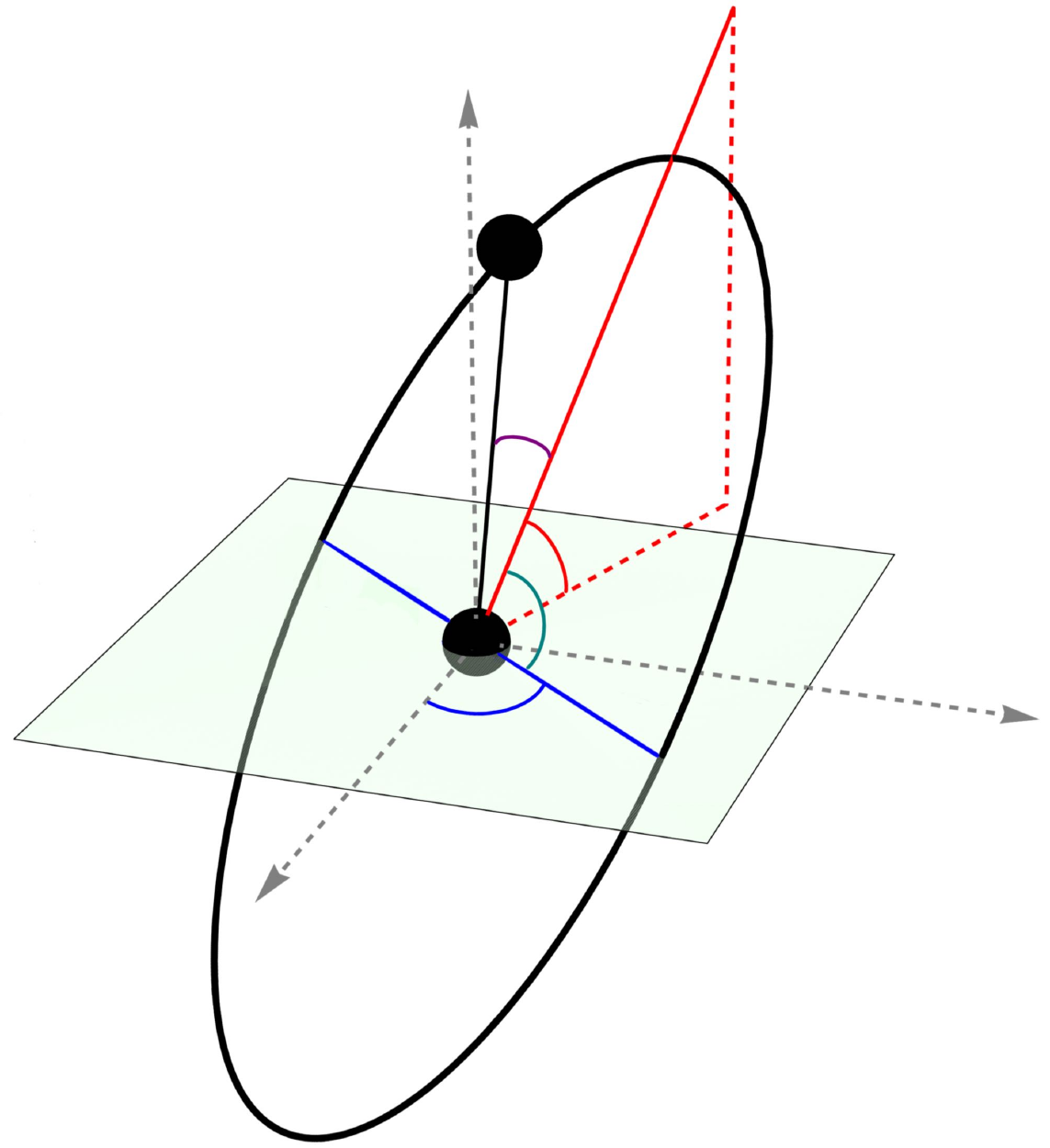}
\begin{picture}(0,0)
    \put(13,157){{\color{red}{$\boldsymbol{I}$}}}
    \put(13,135){{\color{teal}{$\boldsymbol{\gamma}$}}}
    \put(-10,105){{\color{blue}{$\boldsymbol{\theta}$}}}
    \put(3,190){{\color{violet}{$\boldsymbol{\psi}$}}}
    \put(-55,67){\color{gray}{$\boldsymbol{x}$}}
     \put(125,121){\color{gray}{$\boldsymbol{y}$}}
     \put(-19,265){\color{gray}{$\boldsymbol{z}$}}
\end{picture}
\caption{Illustration of orbital parameters. The green plane represents the reference plane while the intersection between the red line and the orbit constitutes the periapsis and the intersection between the blue line and the orbital plane provides the ascending node.}
\label{Fig_orbit}
\end{figure}

One can obtain a generic orientation of the orbit using the Euler angles defined through the  following rotation matrices
\begin{equation}
\begin{split}
R_{\theta}&=\begin{pmatrix}
    \cos\theta &-\sin\theta &0 
    \\
    \sin\theta &\cos\theta &0
    \\
    0&0&1
    \end{pmatrix}
    ~~,~~
    R_{I}=\begin{pmatrix}
    1&0&0
    \\
    0&\cos I &-\sin I
    \\
    0&\sin I&\cos I
    \end{pmatrix}
    ~~,
    \\
    &\qquad\qquad\quad R_\gamma=\begin{pmatrix}
    \cos\gamma &-\sin\gamma&0 
    \\
    \sin\gamma&\cos\gamma&0
    \\
    0&0&1
    \end{pmatrix}~~,
\end{split}
\end{equation}
where the angle $\theta$ is called the \textit{longitude of ascending nodes}, $I$ is the \textit{orbital inclination}, and $\gamma$ is the \textit{argument of the periapsis}. We refer the reader to Fig.~\ref{Fig_orbit} for an illustration of the orbital parameters.
\lbreak

A Newtonian orbit with arbitrary orientation can be obtained by performing a rotation around a reference plane. In our setup, where $m_3\gg (m_1+m_2)$, it is natural to choose this to be the equatorial plane of the Kerr black hole. 
An arbitrary orientation of the inner binary is therefore derived by performing the rotation $\boldsymbol{r}=R_{\theta} R_{I}R_\gamma( r \cos\psi~ \boldsymbol{\hat x}+r \sin\psi ~\boldsymbol{\hat y})$, and by fixing $\psi=0,\pi/2$ one gets
\begin{align}
    \label{eq: uv}
    \nonumber
    \boldsymbol{\hat u}=&(\cos\gamma\cos\theta-\cos I\sin\gamma\sin\theta)~\boldsymbol{\hat x}
    \\ 
    &+(\cos\gamma\sin\theta+\cos I\sin\gamma\cos\theta)~\boldsymbol{\hat y}
    \\ \nonumber
    &+\sin I \sin\gamma~ \boldsymbol{\hat z}~~,
    \\\nonumber
    \boldsymbol{\hat v}=&(-\sin\gamma\cos\theta-\cos I\cos\gamma\sin\theta)~\boldsymbol{\hat x}
    \\
    &+(-\sin\gamma\sin\theta+\cos I\cos\gamma\cos\theta)~\boldsymbol{\hat y}
    \\\nonumber
    &+\sin I \cos\gamma~ \boldsymbol{\hat z}~~.
\end{align}
For later use, we define in addition the \textit{eccentricity vector} as
\begin{equation}
\label{eq: e_vec}
    \boldsymbol{e}=e ~\boldsymbol{\hat u}~~,
\end{equation}
which corresponds to the dimensionless version of the \textit{Laplace–Runge–Lenz} vector \cite{goldstein:mechanics}.
\lbreak

The direction of the angular momentum for the inner binary is readily obtained as $\boldsymbol{\hat L}_{\rm in}=\boldsymbol{\hat u}\times \boldsymbol{\hat v}$, yielding
\begin{equation}
\label{Lin}
    \boldsymbol{\hat L}_{\rm in}=\sin I \sin\theta ~\boldsymbol{\hat x}-\sin I \cos\theta ~\boldsymbol{\hat y}+\cos I ~\boldsymbol{\hat z}~~.
\end{equation}
Since the reference plane in our setup corresponds to the Kerr equatorial plane, $\hat \theta=\pi/2$, and the outer orbit is a fully relativistic equatorial circular geodesic, the magnitude of the outer orbit angular momentum is $\hat L$, as given in \eqref{eq: ELK_Kerr}, and its direction in the distant-star frame is
\begin{equation}
    \boldsymbol{\hat L}_{\rm out}=\boldsymbol{\hat z}~~.
\end{equation}
Notice that, by definition, the inclination angle quantifies the projection of the inner orbit angular momentum on the outer orbit one, $\boldsymbol{\hat L}_{\rm in}\cdot\boldsymbol{\hat L}_{\rm out}=\cos I$.

\subsection{Action-angle variables}
\label{sec:actions}

For any libration of periodic motion, we can introduce action-angle variables to describe the momenta, which is advantageous since they are constants of motion. This is useful for our description of the BBH-SMBH system, as it means that the non-tidal part of the Hamiltonian \eqref{H_newton} is expressed purely via constants of motion.
\lbreak

We shall use the action-angle variables known as \textit{Delauney variables}, with the position given by the three angles
\begin{equation}
(\beta,\gamma,\theta)~~,
\end{equation}
each periodic with period $2\pi$, as well as
 the corresponding action-angle variables
\begin{equation}
\label{eq: J_def1}
\begin{split}
    J_\beta=\mu&\sqrt{G M a}~~,~~ J_\gamma=\mu\sqrt{G M a(1-e^2)}~~,
    \\
    & J_\theta=\mu\sqrt{G M a(1-e^2)} \cos I~~,
\end{split}
\end{equation}
with $M$ and $\mu$ defined in Eq.~\eqref{eq: comframe} as the total mass and the reduced mass of the BBH system respectively.
\lbreak

It is worth noting that the Delauney action variables are related to the magnitude and the orientation of the angular momentum of the inner binary with respect to the reference plane. In particular
\begin{equation}
\label{eq: J_def2}
    J_\gamma=|\boldsymbol{L}_{\rm in}|~~,~~J_\theta=\boldsymbol{L}_{\rm in}\cdot \boldsymbol{\hat{L}}_{\rm out}~~.
\end{equation}
The total Hamiltonian of the inner binary \eqref{HBini} in the distant-star frame of reference  can therefore be simply expressed as
\begin{equation}
\label{eq: H_Del_DS}
    \mathcal{H}=-\left(\frac{G M}{J_\beta}\right)^2+\Omega_{\rm g} J_\theta
    +\mathcal{H}_{\rm q}~~.
\end{equation}
and the quadrupole tidal part is expressed in terms of Euler angles according to%
\begin{equation}
\begin{split}
\label{theHq_DS}
    \mathcal{H}_{\rm q}=\frac{\mu}{2}&\frac{Gm_3r^2}{\hat r^3}\bigg[1+3\frac{K}{\hat r^2}\sin^2(\gamma+\psi)\sin^2 I\\
    &-3\left(1+\frac{K}{\hat r^2}\right)\big(\cos(\hat{\phi}-\theta)\cos(\gamma+\psi)
    \\
    &+\sin(\hat{\phi}-\theta)\sin(\gamma+\psi)\cos I\big)^2\bigg]~~.
\end{split}
\end{equation}
We notice that the non-tidal part of \eqref{eq: H_Del_DS} indeed is given by an action-angle variable.

\subsection{Action-angle variables for Marck's frame}
\label{sec:marck_variables}

For use below, we also introduce the action-angle variables for the Hamiltonian in Marck's frame of reference given by Eqs.~\eqref{H_newton} and \eqref{eq: tidal_pot}.

It follows from Section \ref{sec:gyroscope} that the longitude of ascending nodes for Marck's frame of reference is
\begin{equation}
\theta' = \theta - \Omega_{\rm g} \tau~~,
\end{equation}
in terms of the corresponding angle $\theta$ of the distant-star frame. We can also write this as $\theta'-\Psi=\theta-\hat{\phi}$.
Instead, the other angles $\beta$ and $\gamma$ remain the same. 
\lbreak

The Hamiltonian along with the action-angle variables conjugate to the angles $(\beta,\gamma,\theta')$ are most efficiently found using a canonical transformation of the second type \cite{goldstein:mechanics}. 
This reveals that the momenta $(J_\beta,J_\gamma,J_\theta)$ given in the distant-star frame \eqref{eq: J_def1} are the same for Marck's frame, for it to be a canonical transformation. In detail, we have the
generating function
\begin{equation}
    F_2(q,P,\tau)= \beta~J_\beta + \gamma ~J_\gamma + (\theta-\Omega_{\rm g} \tau)~J_\theta~~,
\end{equation}
with $(q,p)$  and $(Q,P)$ being the distant-star and the Marck phase space variables, respectively, and with the identifications
\begin{equation}
\begin{split}
    &q_i = (\beta,\gamma,\theta) \spa Q_i = (\beta,\gamma,\theta')\,, \\  &p_i = P_i = (J_\beta,J_\gamma,J_\theta)  \,.
    \end{split}
\end{equation}
This gives $p_i = \partial F_2/ q_i$ and $Q_i = \partial F_2/ \partial P_i$ as needed.
The transformed Hamiltonian therefore becomes 
\begin{equation} 
\label{transformed_Ham}
 H
 = \mathcal{H}
  +\partial_\tau F_2 =  \mathcal{H}  - \Omega_{\rm g} J_\theta~.
\end{equation}
From this we get the Hamiltonian in Marck's frame as
\begin{equation}
\label{eq: H_Del_Marck}
    H=-\left(\frac{G M}{J_\beta}\right)^2
    +H_{\rm q}~~.
\end{equation}
with quadrupole tidal part 
\begin{equation}
\begin{split}
\label{theHq_Marck}
    H_{\rm q}=\frac{\mu}{2}\frac{Gm_3r^2}{\hat r^3}\bigg[&1+3\frac{K}{\hat r^2}\sin^2(\gamma+\psi)\sin^2 I\\
    &-3\left(1+\frac{K}{\hat r^2}\right)\big(\cos(\Psi-\theta' )\cos(\gamma+\psi)
    \\
    &+\sin(\Psi-\theta')\sin(\gamma+\psi)\cos I\big)^2\bigg]~~.
\end{split}
\end{equation}
One can easily check that if one introduces Delauney/action-angle variables directly for Marck's frame Hamiltonian given by Eqs.~\eqref{H_newton} and \eqref{eq: tidal_pot} one would get the same result as above.

\section{Secular Hamiltonian}
\label{sec: <H>}

In this section, we obtain the secular dynamics of the BBH-SMBH system. The secular dynamics describes the system at timescales that are larger than both the inner and outer orbit periods, {\textit i.e.} both the time scale of internal motion within the BBH as well as the time scale of the geodesic motion of the BBH around the Kerr black hole. 
\lbreak

The secular Hamiltonian is obtained by taking the average both over the inner orbit motion as well as the outer orbit motion. As explained above, these two motions can be separated to leading order in our regime \eqref{SEPtidal}, in that the outer orbit motion corresponds to the center of mass of the binary system moving on a circular geodesic in the equatorial plane of the supermassive Kerr black hole. 
\lbreak

The inner binary is described by a Newtonian elliptic motion, which is perturbed by tidal forces.\footnote{In Section \ref{sec: GW} we shall include the 1PN effect of the periastron precession as well as the leading GW radiation-reaction effect.} To take the average, we need an angle that grows uniformly with time in the elliptic motion. This is provided by the 
mean anomaly $\beta$ defined in Eq.~\eqref{beta_psi}. However, since the tidal part of the Hamiltonian \eqref{theHq_DS} is a function of the true anomaly $\psi$ instead, we translate $(2\pi)^{-1} \int_0^{2\pi} d\beta$ into 
\begin{equation}
  \frac{1}{2\pi}  \int_0^{2\pi}d\psi\frac{(1-e^2)^{3/2}}{(1+e\cos\psi)^2}~~,
\end{equation}
since this follows from Eq.~\eqref{beta_psi}.

For the outer binary, the averaging procedure is more subtle, reflecting the fact that general relativistic effects play a role. 
In terms of the action-angle variables introduced in Section \ref{sec: euler_angles_and_action_angle_variables}, we have found the Hamiltonian in both the non-inertial distant-star frame with Eqs.~\eqref{eq: H_Del_DS}-\eqref{theHq_DS} as well as in the inertial Marck's frame with Eqs.~\eqref{eq: H_Del_Marck}-\eqref{theHq_Marck}.
\lbreak

We begin by considering the outer orbit average in the distant-star frame since this bears the closest resemblance to the Newtonian case (see for example \cite{Randall:2018qna}) and therefore is more intuitive. 
Indeed, considering the Hamiltonian~\eqref{eq: H_Del_DS}-\eqref{theHq_DS}, we notice that it is periodic in the angle $\hat{\phi}$ with period $2\pi$. 
This periodicity is precisely associated with one outer orbit cycle of motion. Moreover, the angle grows linearly with proper time, as one can infer from Eq.~\eqref{hatphi}, making it the relativistic analog of the outer orbit angle,  that one for instance uses in \cite{Randall:2018qna}.
Therefore, the outer orbit average in this frame is simply performed as $(2\pi)^{-1} \int_0^{2\pi} d\hat{\phi}$. 
\lbreak

The secular Hamiltonian is thus computed as the following double-average of the Hamiltonian \eqref{eq: H_Del_DS} in the distant-star frame
\begin{equation}
\label{eq: H_secular_DS}
    \langle{\mathcal{H}}\rangle=-\left(\frac{G M}{J_\beta}\right)^2
    +\Omega_{\rm g} J_\theta
    +\langle{\mathcal{H}_{\rm q}}\rangle~~,
\end{equation}
with
\begin{equation}
\label{eq: H_secular_DS_average}
    \langle \mathcal{H}_{\rm q}\rangle\equiv\frac{1}{(2\pi)^2}\int_0^{2\pi}d\hat{\phi} \int_0^{2\pi}d\psi\frac{(1-e^2)^{3/2}}{(1+e\cos\psi)^2}\mathcal{H}_{\rm q}~~,
\end{equation}
which explicitly reads 
\begin{equation}\label{eq: H_secular_extra}
\begin{split}
    & \langle \mathcal{H}_{\rm q}\rangle=-\Omega_{\mbox{\tiny ZLK}}^{(\rm GR)} J_\gamma\left(\mathcal{W}+\frac{5}{3}\right)~~,
    \\
    &\mathcal{W}=(1-e^2)(\cos^2I-2)-5e^2\sin^2I\sin^2\gamma~~,
    \\
    &\Omega_{\mbox{\tiny ZLK}}^{(\rm GR)}=\Omega_{\mbox{\tiny ZLK}}^{(\rm N)}\left(1+3\frac{ K}{\hat r^2}\right)~,\\
    &\Omega_{\mbox{\tiny ZLK}}^{(\rm N)}=\frac{3}{8 J_\gamma}\left(\frac{G m_3\mu}{\hat r}\right)\left(\frac{ a }{\hat{r}}\right)^2~~,
    ~~
\end{split}
\end{equation}
where the subscript ${\mbox{\tiny ZLK}}$ refers to the ZLK effect which will be extensively discussed in the remaining Sec.~\ref{sec: relativistic_effects_KL} and in Sec.~\ref{sec: GW}.
\footnote{\label{foot8} Note that \cite{Maeda:2023tao} has an expression for the secular Hamiltonian of the binary system, for the case of SMBH without spin. However, it seems to include an extra term that is quadratic in $\Omega_{\rm g}$ which is at odds with the fact that one cannot observe the gyroscope precession in the inertial frame of reference of Marck, as a consequence of the equivalence principle.}
\lbreak

The expressions above make manifest the fact that the GR effects associated with the Kerr perturber enter the dynamics through the overall prefactor $\Omega_{\mbox{\tiny ZLK}}^{(\rm GR)}$ in the averaged tidal Hamiltonian \eqref{eq: H_secular_extra}. It is immediate to see from~\eqref{eq: H_secular_extra} that in the weak field regime $\hat{r}\to\infty$ one recovers the Newtonian secular Hamiltonian of Ref.~\cite{Randall:2018qna}. These GR effects are completely accounted for by the term in $\Omega_{\mbox{\tiny ZLK}}^{(\rm GR)}$ proportional to the Carter constant $K$.
\lbreak

However, there are other GR effects as well, some of which are not immediately apparent from the above secular Hamiltonian of Eqs.~\eqref{eq: H_secular_DS} and \eqref{eq: H_secular_extra}. 
The most obvious one is the gyroscope precession of the binary system introduced in Section \ref{sec:gyroscope}, here arising from the term $\Omega_{\rm g} J_\theta$ in the total Hamiltonian \eqref{eq: H_secular_DS}.
Another less explicit effect is the time dilation of the proper time used above, relative to the asymptotic time $\hat{t}$. This we shall include later in Section \ref{sec: GW}. 
Both of these effects are related to how an asymptotic observer will view the binary system, {\textit i.e.}~the global point of view, rather than the local point of view. Furthermore, in Section \ref{sec: GW} we shall add further relativistic effects to the binary dynamics, to describe gravitational backreaction due to gravitational waves.
\lbreak

So far we have considered  the averaging procedure only in the distant-star frame.
It is important to check that one can obtain the same secular average in Marck's frame of reference. We notice immediately that the Hamiltonian \eqref{eq: H_Del_Marck}-\eqref{theHq_Marck} is periodic in $\Psi$ with period $2\pi$. 
Also, the angle grows linearly with time $\Psi = \Omega_{\Psi} \tau$. Therefore, we conclude that, in Marck's frame, one should compute the outer orbit average as $(2\pi)^{-1} \int_0^{2\pi} d\Psi$. Explicitly, 
\begin{equation}
\label{eq: H_secular_Marck}
    \langle{H}\rangle=-\left(\frac{G M}{J_\beta}\right)^2
    +\langle{H_{\rm q}}\rangle~~.
\end{equation}
with
\begin{equation}
\label{eq: H_secular_Marck_average}
    \langle{H}_{\rm q}\rangle\equiv\frac{1}{(2\pi)^2}\int_0^{2\pi}d\Psi \int_0^{2\pi}d\psi\frac{(1-e^2)^{3/2}}{(1+e\cos\psi)^2}H_{\rm q}~~,
\end{equation}
It is now straightforward to see that
double-average over the tidal part of the Hamiltonian in the two different frames agree
\begin{equation}
\langle{H}_{\rm q}\rangle = \langle \mathcal{H}_{\rm q}\rangle~~.
\end{equation}
This means the only difference between the secular Hamiltonians in the two frames is the constant term $\Omega_{\rm g} J_\theta$, accounting for the fictitious forces. 
\lbreak

However, as explained in Section \ref{sec:gyroscope}, the gyroscope precession essentially measures the difference between $\hat{\phi}$ and $\Psi$ when they have gone through one cycle in the outer orbit motion. So how can the outer orbit average over the tidal contribution give the same result in the two different frames, as we are averaging over two different angles? The answer lies in the formula $\hat{\phi}-\theta=\Psi-\theta'$. In the distant-star frame one should keep fixed $\theta$ in taking the average, as this angle is fixed during the motion. But, for Marck's frame, it is instead the angle $\theta'$ that one should keep fixed. Thus, the reason that the outer orbit averages give the same result in the two frames is that the difference between $\theta$ and $\theta'$ precisely accounts for the difference $\hat{\phi}-\Psi$, which is the gyroscope precession. \\
Notice that the canonical transformation detailed in Sec.~\ref{sec:marck_variables} can also be directly used to relate the secular Hamiltonian in the distant-star frame with the secular Hamiltonian in Marck's frame. Since neither depend on the respective longitude of ascending nodes angle ($\theta$ and $\theta'$), this transformation simply relates the secular Hamiltonians as $\langle H\rangle = \langle \mathcal{H}\rangle-\Omega_{\rm g} J_\theta$.

\section{ZLK mechanism in a strong GR background}
\label{sec: relativistic_effects_KL}

In this section, we apply the result for the secular Hamiltonian in the distant-star reference frame, as derived in the previous section, to study the long timescale dynamics of the BBH system moving on an equatorial circular geodesic of the external Kerr SMBH. 
\lbreak

An important result of Sec.~\ref{sec: first_look} is that we can
quantify to what extent the ZLK frequency departs from its Newtonian value when one takes into account strong gravity effects associated to the general relativistic description that we adopt for the outer orbit. 
\lbreak

Moreover, we derive in Section \ref{sec: evo_eqs} the equations of motion for the inner orbital parameters, from which one can study the evolution of the ZLK mechanism. This is used in Section \ref{sec: rel_effects} to compare the weak-gravity limit of our results to the PN corrections found in the literature. 
\lbreak

In Section \ref{sec: GW} we build on the results of this section by refining the equations of motion for the inner orbital parameters found in Section \ref{sec: rel_effects} to include the periastron precession and GW emission. This is used to study the ZLK mechanism and its influence on the binary merger time.

\subsection{ZLK frequency in the vicinity of SMBH}
\label{sec: first_look}

From the distant-star frame secular Hamiltonian, given by Eqs.~\eqref{eq: H_secular_DS} and \eqref{eq: H_secular_extra}, it is immediate to observe that two main effects govern the secular dynamics of the BBH system: the ZLK mechanism, which manifests due to the tidal interaction with the external SMBH, and the gyroscope precession, which is present in the distant-star frame of reference whenever a general relativistic description for the SMBH is adopted.
\lbreak

We begin this section by discussing the ZLK mechanism
which in the distant-star secular Hamiltonian is modeled by the quadrupole tidal contribution $\langle \mathcal{H}_q \rangle$ in Eq.~\eqref{eq: H_secular_extra}. A well-known result in the literature concerning the ZLK mechanism \cite{1976CeMec..13..471L,1962AJ.....67..591K, doi:10.1146/annurev-astro-081915-023315} is that the set of parameters characterizing the outer orbit only enters in $\langle \mathcal{H}_q \rangle $ through the frequency of the eccentricity/inclination oscillations. 
We observe from \eqref{eq: H_secular_extra} that this remains true in our case, as all the information concerning the outer orbit enter through the frequency $\Omega_{\mbox{\tiny ZLK}}^{(\rm GR)}$. Thus, all new tidal force effects that arise from an exact metric description of the SMBH as a Kerr black hole enters through this frequency.
\lbreak

Therefore, the main aim of the following is to show that BBH systems close enough to an external SMBH to probe the strong gravity regime can manifest substantial deviations in the frequency of the ZLK oscillations, compared to 
the frequency one gets from employing a Newtonian point particle approximation.
\lbreak

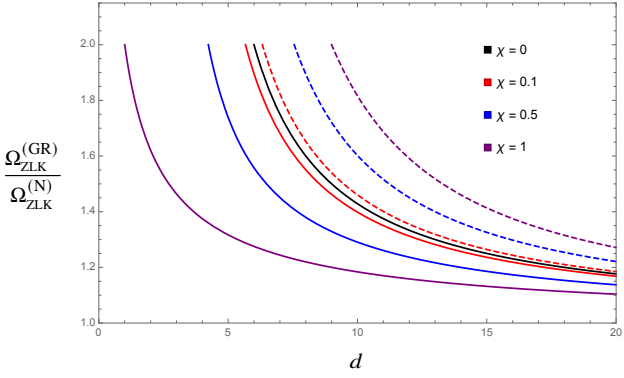
\begin{figure}
    \centering
    \includegraphics[width=0.40\textwidth]{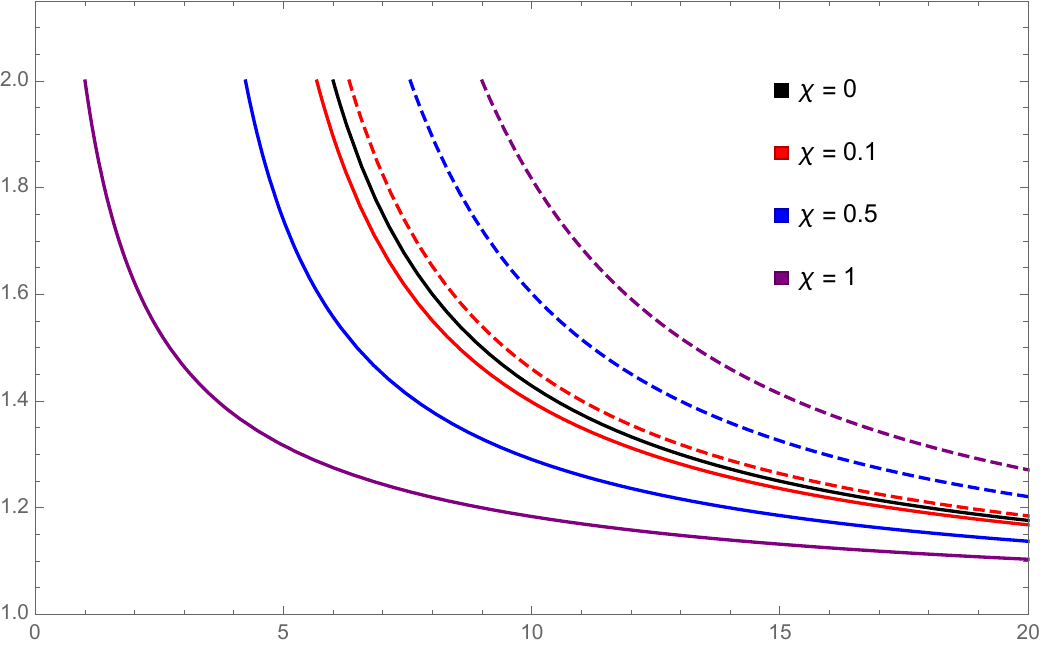}
     \\
    \begin{picture}(0,0)
        \put(0,0){$d$}
        \put(-130,70){{{$\dfrac{\Omega_{\mbox{\tiny ZLK}}^{(\rm GR)}}{\Omega_{\mbox{\tiny ZLK}}^{(\rm N)}}$}}}
    \end{picture}
    \caption{Diagram of the ratio $\Omega_{\mbox{\tiny ZLK}}^{(\rm GR)}/\Omega_{\mbox{\tiny ZLK}}^{(\rm N)}$ versus the distance $d$ from the SMBH.
    Different colors label various values of the black hole spin $\chi$ where the solid and dashed lines represent co-rotating ($\sigma=+1$) and counter-rotating ($\sigma=-1$) orbits, respectively. Each curve terminates at the ISCO.}
    \label{fig: Omega_Ratio}
\end{figure}

We begin by comparing $\Omega_{\mbox{\tiny ZLK}}^{(\rm GR)}$ to the 
Newtonian frequency $\Omega_{\mbox{\tiny ZLK}}^{(\rm N)}$, {\sl i.e.} the frequency that would have resulted from Newtonian quadrupole tidal forces induced by a particle of mass $m_3$. We find
\begin{equation}
\label{eq: Omega_KL}
\dfrac{\Omega_{\mbox{\tiny ZLK}}^{(\rm GR)}}{\Omega_{\mbox{\tiny ZLK}}^{(\rm N)}}
= \frac{1+\frac{3\chi^2}{d^2}-\frac{4\sigma \chi}{d^{3/2}}}{1-\frac{3}{d} +\frac{2\sigma\chi}{d^{3/2}}}~~,
\end{equation}
where we defined for convenience the dimensionless radius $d$ for the equatorial circular orbit as
\begin{equation}
d= \hat{r} \frac{c^2}{Gm_3}~~.
\label{eq:d}
\end{equation}
Remarkably, at the ISCO $\hat{r}=\hat{r}^\sigma_{\rm ISCO}$ the ratio \eqref{eq: Omega_KL} takes the universal value%
\footnote{One can derive this using $K=\tfrac{1}{3}\hat r_{\rm ISCO}^2$ at the ISCO, see Ref.~\cite{Camilloni:2023rra}.}
\begin{equation}
\label{eq: Omega_ISCO}
    \Omega_{\mbox{\tiny ZLK}}^{(\rm GR)}=2\Omega_{\mbox{\tiny ZLK}}^{(\rm N)}~~.
\end{equation}
This result will be highly important in Section \ref{sec: GW} where we consider the evolution of the BBH-SMBH system in detail. 
One can check that \eqref{eq: Omega_ISCO} gives the maximal value of the ratio \eqref{eq: Omega_KL} that the binary system can attain. Instead, for large $d$ the ratio goes to one. Both of these statements are illustrated in Fig.~\ref{fig: Omega_Ratio}.
The result \eqref{eq: Omega_ISCO} shows that one has an order one difference between the weak-field Newtonian result $\Omega_{\mbox{\tiny ZLK}}^{(\rm N)}$ and our novel strong field  result $\Omega_{\mbox{\tiny ZLK}}^{(\rm GR)}$ when close to the SMBH.

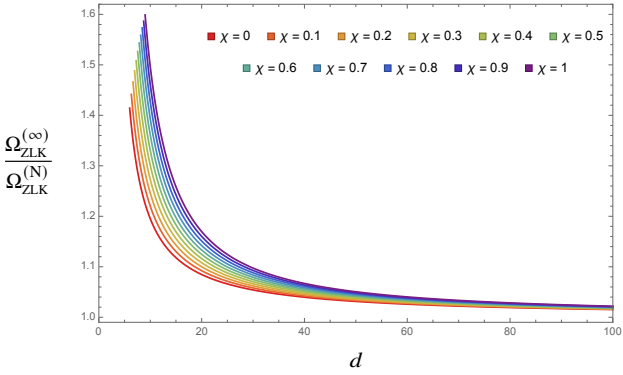
\begin{figure}
    \centering
    \includegraphics[width=0.40\textwidth]{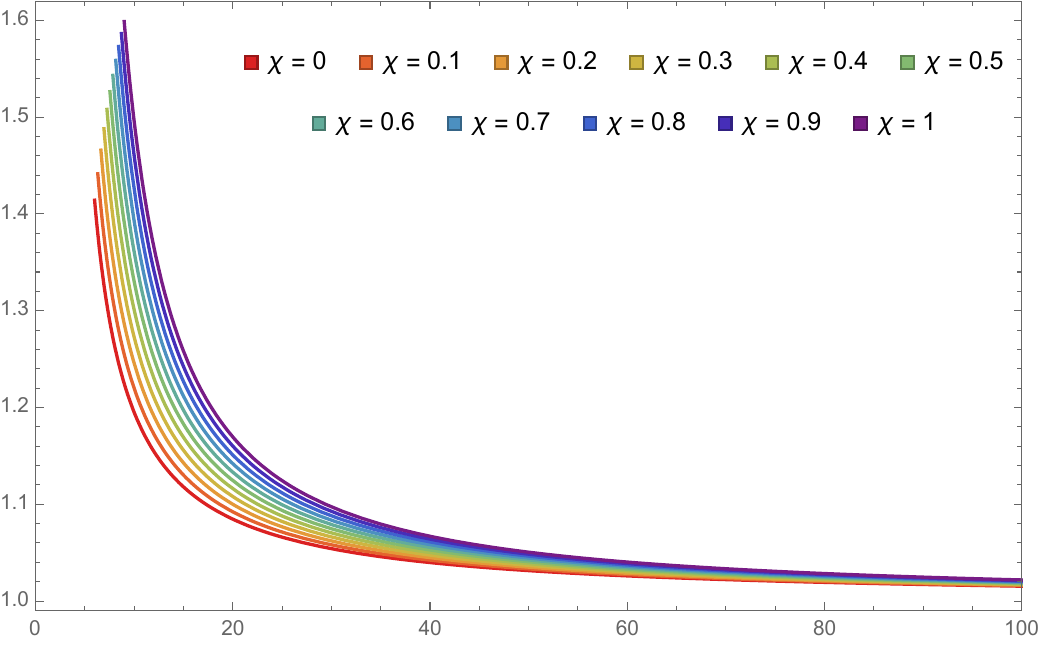}
     \\
    \begin{picture}(0,0)
        \put(0,0){$d$}
        \put(-130,75){{{$\dfrac{\Omega_{\mbox{\tiny ZLK}}^{(\infty)}}{\Omega_{\mbox{\tiny ZLK}}^{(\rm N)}}$}}}
    \end{picture}
    \caption{The ratio $\Omega_{\mbox{\tiny ZLK}}^{(\infty)}/{\Omega_{\mbox{\tiny ZLK}}^{(\rm N)}}$ is plotted in the counter-rotating case $\sigma=-1$ as a function of the dimensionless radius $d$ for several values of the spin.
    The figure shows that the maximum value for the ratio  is always at the ISCO.}
    \label{fig: Omega_Ratio_asympt_counter}
\end{figure}

However, it is important to note here that the frequency $\Omega_{\mbox{\tiny ZLK}}^{(\rm GR)}$ is measured with respect to the proper time of the BBH orbit. Thus, this is not the frequency that an asymptotic observer would measure. To find the corresponding asymptotic ZLK frequency we need to incorporate the redshift factor as follows
\begin{equation}
\label{Omegainf}
    \Omega_{\mbox{\tiny ZLK}}^{(\infty)} = \frac{1}{u^t} \Omega_{\mbox{\tiny ZLK}}^{(\rm GR)} ~~
\end{equation}
where the redshift factor can be written as
\begin{equation}
\label{eq: ut}
    u^t=
    \frac{\Omega_{\hat \phi}}{\Omega}
= \frac{1+\frac{\sigma\chi}{d^{3/2}}}{\sqrt{1-\frac{3}{d} +\frac{2\sigma\chi}{d^{3/2}}}} ~~,
\end{equation}
which one can check is always greater than 1, and it is a decreasing function of $\sigma \chi$ for fixed $d$.
Using \eqref{eq: ut} we get the following ratio between the asymptotically measured ZLK frequency $\Omega_{\mbox{\tiny ZLK}}^{(\infty)}$, which now takes into account all the GR effects, and the corresponding Newtonian frequency $\Omega_{\mbox{\tiny ZLK}}^{(\rm N)}$
\begin{equation}
\label{eq: Omega_KL_asympt}
\dfrac{\Omega_{\mbox{\tiny ZLK}}^{(\infty)}}{\Omega_{\mbox{\tiny ZLK}}^{(\rm N)}}
= \frac{1+\frac{3\chi^2}{d^2}-\frac{4\sigma \chi}{d^{3/2}}}{\left(1+\frac{\sigma\chi}{d^{3/2}}\right)\sqrt{1-\frac{3}{d} +\frac{2\sigma\chi}{d^{3/2}}}} ~~.
\end{equation}
The inclusion of the redshift factor gives a more refined difference in the ratio of the frequencies. 
One finds that in the counter-rotating case $\sigma=-1$ the 
maximal value of the ratio \eqref{eq: Omega_KL_asympt} is at the ISCO, as illustrated in Fig.~\ref{fig: Omega_Ratio_asympt_counter}. 
For the co-rotating case $\sigma=1$, the same is true for the range $0 \leq \chi \leq 0.69$. However, as illustrated in Fig.~\ref{fig: Omega_Ratio_asympt_zoom}, this behavior starts changing in the range $0.69 \leq \chi \leq 0.7$, so that for $\chi \geq 0.7$ the maximal value of the ratio \eqref{eq: Omega_KL_asympt} is no longer reached at the ISCO. We see from Fig.~\ref{fig: Omega_Ratio_asympt} that for $\chi \geq 0.75$ it is instead the minimal value that one reaches at the ISCO.

\begin{figure}
    \centering
    \includegraphics[width=0.40\textwidth]{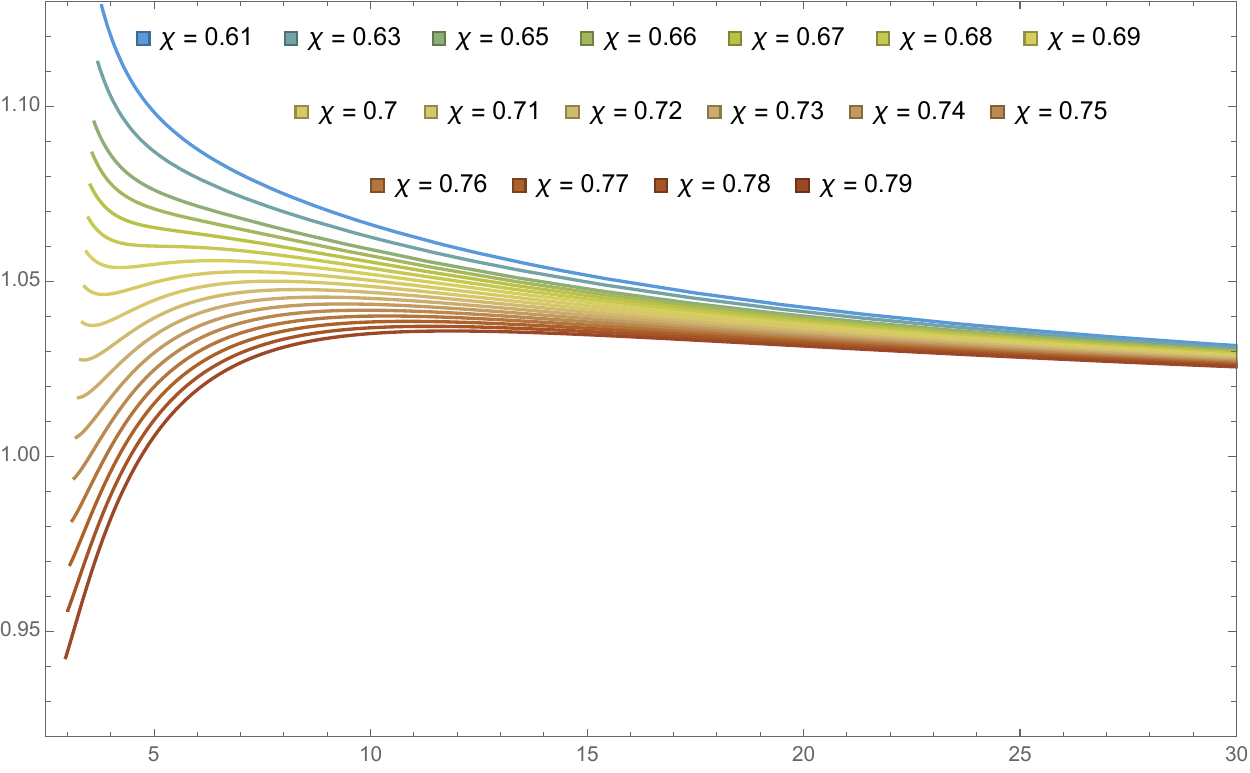}
     \\
    \begin{picture}(0,0)
        \put(0,0){$d$}
        \put(-130,75){{{$\dfrac{\Omega_{\mbox{\tiny ZLK}}^{(\infty)}}{\Omega_{\mbox{\tiny ZLK}}^{(\rm N)}}$}}}
    \end{picture}
    \caption{The ratio $\Omega_{\mbox{\tiny ZLK}}^{(\infty)}/{\Omega_{\mbox{\tiny ZLK}}^{(\rm N)}}$ is plotted in the co-rotating case $\sigma=1$ as function of the dimensionless radius $d$ for the range $0.61 \leq \chi \leq 0.79$.}
    \label{fig: Omega_Ratio_asympt_zoom}
\end{figure}
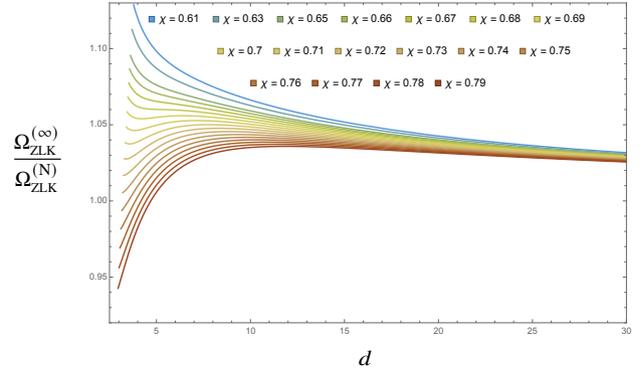
\begin{figure}
    \centering
    \includegraphics[width=0.40\textwidth]{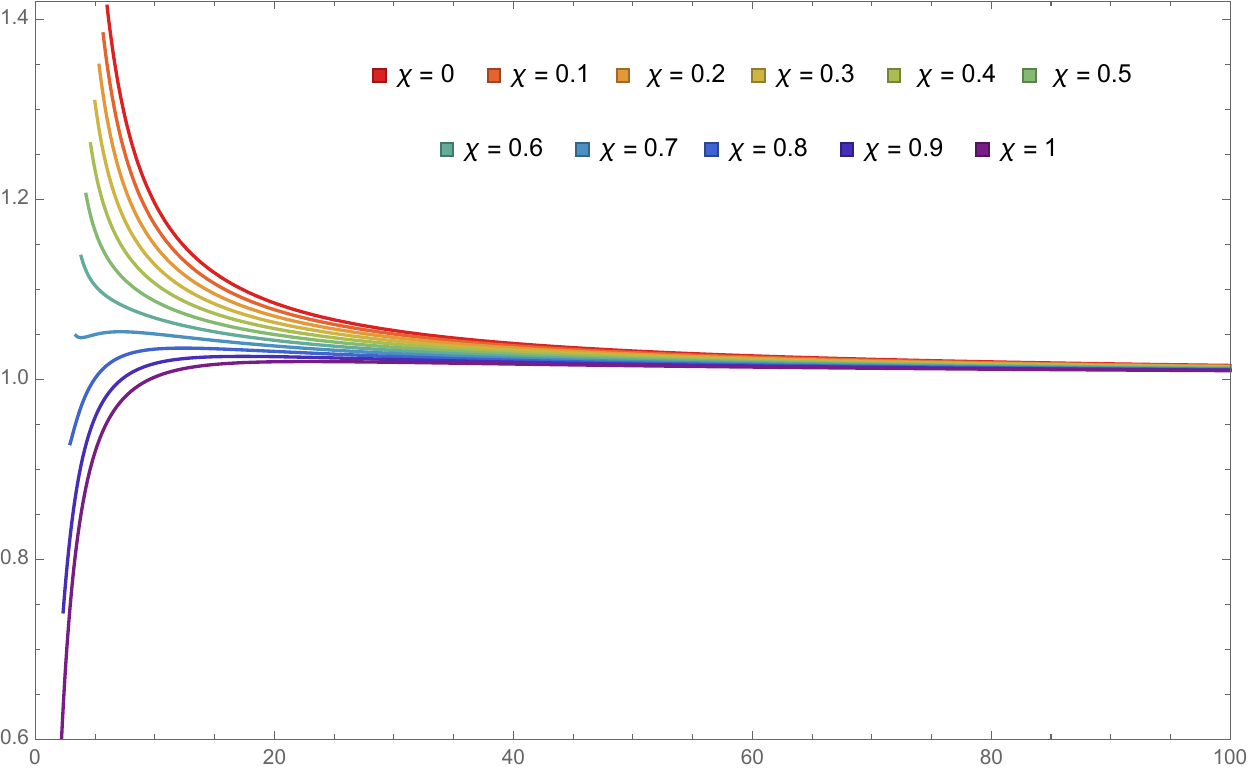}
     \\
    \begin{picture}(0,0)
        \put(0,0){$d$}
        \put(-130,75){{{$\dfrac{\Omega_{\mbox{\tiny ZLK}}^{(\infty)}}{\Omega_{\mbox{\tiny ZLK}}^{(\rm N)}}$}}}
    \end{picture}
    \caption{The ratio $\Omega_{\mbox{\tiny ZLK}}^{(\infty)}/{\Omega_{\mbox{\tiny ZLK}}^{(\rm N)}}$ is plotted in the co-rotating case $\sigma=1$ as a function of the dimensionless radius $d$ for several values of the spin.}
    \label{fig: Omega_Ratio_asympt}
\end{figure}
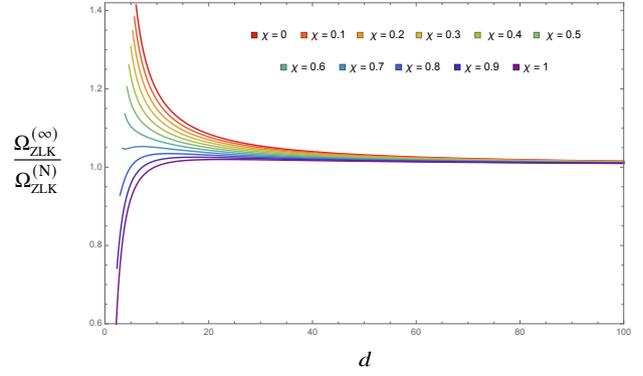
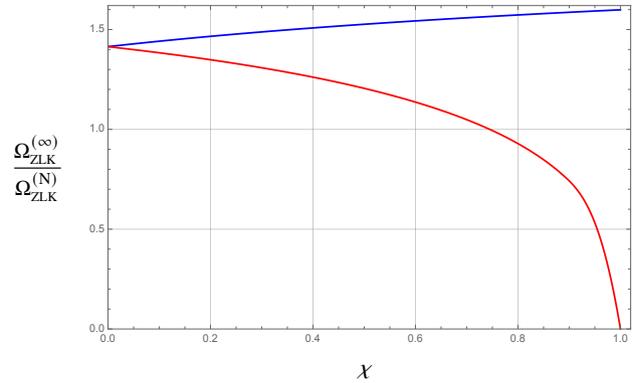
\begin{figure}
    \centering
    \includegraphics[width=0.40\textwidth]{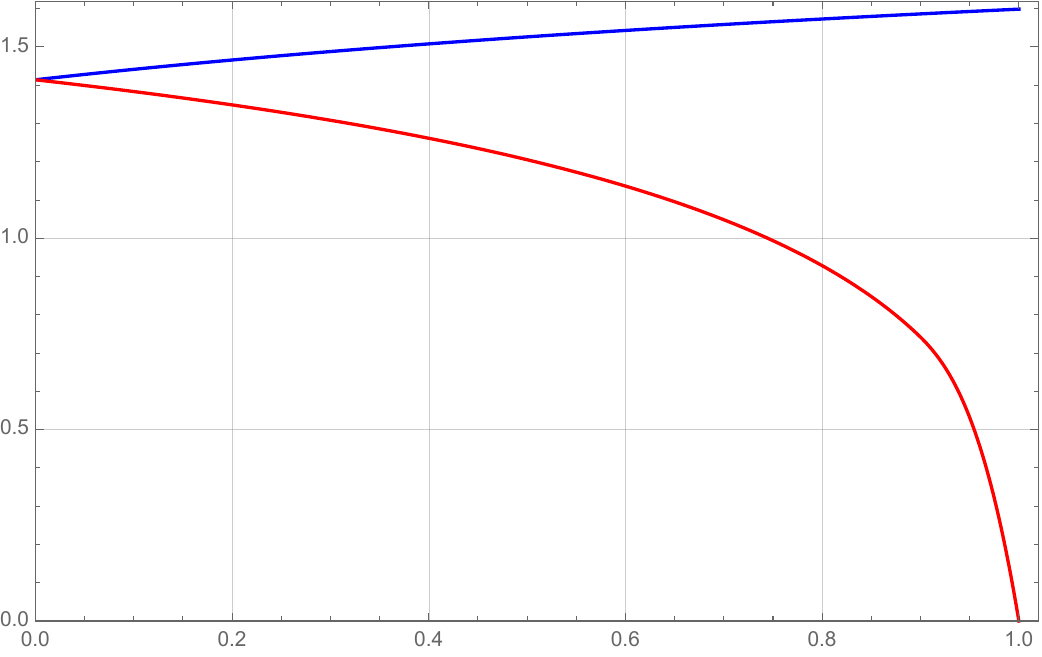}
     \\
    \begin{picture}(0,0)
        \put(0,0){$\chi$}
        \put(-130,75){{{$\dfrac{\Omega_{\mbox{\tiny ZLK}}^{(\infty)}}{\Omega_{\mbox{\tiny ZLK}}^{(\rm N)}}$}}}
    \end{picture}
    \caption{Here the binary system is always placed at the ISCO. The blue line describes the counter-rotating case while the red line describes the co-rotating case.}
    \label{fig: Omega_Ratio_ISCO_asympt_3}
\end{figure}

For all three Figs.~ \ref{fig: Omega_Ratio_asympt_counter}, \ref{fig: Omega_Ratio_asympt_zoom}, and  \ref{fig: Omega_Ratio_asympt} we note that the ratio \eqref{eq: Omega_KL_asympt} approaches 1 for $d$ going to infinity, as one would expect.
Finally, we have also plotted the value of the ratio \eqref{eq: Omega_KL_asympt} at the ISCO in Fig.~\ref{fig: Omega_Ratio_ISCO_asympt_3}.
\lbreak

The above results for $\Omega_{\mbox{\tiny ZLK}}^{(\infty)}/\Omega_{\mbox{\tiny ZLK}}^{(\rm N)}$ show that the GR effects that arise from being in close vicinity to the SMBH are highly significant. As mentioned in the Introduction, this is particularly relevant in the case of bound systems of BBHs situated in the GC.
Indeed, we see the importance of including strong-gravity effects as they significantly alter the frequency, and therefore the timescale, of the ZLK oscillations. One can also see that the spin of the SMBH, as modeled by a Kerr black hole, can significantly alter the dynamics.

\subsection{Evolution equations for ZLK mechanism}
\label{sec: evo_eqs}

Using the secular Hamiltonian~\eqref{eq: H_secular_extra} we can now derive the evolution equations for the orbital variables describing the inner BBH system. It is possible to derive an evolution equation for the orbital inclination $I$ by exploiting the fact that $J_\theta=J_\gamma\cos I$, Eq.~\eqref{eq: J_def1}, which yields 
\begin{equation}
\label{eq: dIdt}
    \frac{dI}{d\tau}=\frac{1}{J_{\gamma}\sin I}\left(\frac{d J_{\gamma}}{d\tau}\cos I-\frac{d J_{\theta}}{d\tau}\right)~~.
\end{equation}
This general equation will be useful when including the loss of angular momentum associated with the emission of GWs, which we postpone to Sec. \ref{sec: GW}. In this section instead,  we only focus on the effect of tidal deformations resulting in the ZLK mechanism. For the secular Hamiltonian \eqref{eq: H_secular_extra}  the Euler angle $\theta$ is a cyclic variable so that its conjugate momentum $J_\theta$ is a constant of motion, $dJ_{\theta}/d\tau=-\partial_\theta \langle{\mathcal{H}}\rangle=0$, and the last term in \eqref{eq: dIdt} does not contribute if we ignore the emission of GWs.
\lbreak

Similarly, the equation of motion for the eccentricity of the inner binary follows from the definition of the Delauney variable $J_\gamma$ in Eq.~\eqref{eq: J_def1}. In the following we use the fact that no variation of the semi-major axis $a$ exists in the absence of GW emission, so that
\begin{equation}
\label{dedt}
    \frac{d e}{d\tau}=\frac{d J_{\gamma}}{d\tau}\left(\frac{d J_{\gamma}}{de}\right)^{-1}=-\frac{1-e^2}{e}\frac{1}{J_{\gamma}}\frac{d J_{\gamma}}{d\tau}~~.
\end{equation}
Therefore, the evolution equations for the orbital elements can be derived from the equation of motion of $J_{\gamma}$. This follows from Hamilton's equations and only involves the tidal Hamiltonian
\begin{equation}
    \label{eq: Jgamma1dot}
    \begin{split}
    &\frac{d J_{\gamma}}{d\tau}=-\frac{\partial \langle\mathcal{H}\rangle}{\partial \gamma}=-5\Omega_{\mbox{\tiny ZLK}}^{(\rm GR)}J_\gamma e^2\sin^2I \sin2\gamma~~.
    \end{split}
\end{equation}
From the relations \eqref{eq: dIdt}, \eqref{dedt}, and \eqref{eq: Jgamma1dot} one can derive  the ZLK contributions to the evolution equations for the orbital inclination $I$ and for the eccentricity $e$ 
\begin{align}
\label{eq: I_evolution}
        &\frac{d I}{d\tau}=-\frac{5}{2}\Omega_{\mbox{\tiny ZLK}}^{(\rm GR)} e^2\sin 2I \sin2\gamma~~,
        \\
\label{eq: e_evolution}
        &\frac{d e}{d\tau}=5\,\Omega_{\mbox{\tiny ZLK}}^{(\rm GR)} e(1-e^2)\sin^2I \sin2\gamma~~.
\end{align}
From these equations of motion, it is immediate to notice that the stationary points for minimum and maximum inclination and eccentricity correspond to $\gamma=0,\pi/2$. As we mentioned earlier, as long as the gravitational backreaction is neglected, $J_\theta=\boldsymbol{L}_{\rm in}\cdot \boldsymbol{\hat L}_{\rm out}= \mu\sqrt{G M a (1-e^2)}\cos I$ is conserved, meaning that the orbital inclination has a maximum $I_{\rm max}$ when the eccentricity is minimal $e_{\rm min}$, and viceversa. In particular, by computing the second derivative in \eqref{eq: I_evolution} and \eqref{eq: e_evolution}, it is easy to show that the pair $(I_{\rm max},e_{\rm min})$ occur for $\gamma=0$, whereas one has $(I_{\rm min},e_{\rm max})$ for $\gamma=\pi/2$.
\lbreak

The evolution equation for the longitude of ascending nodes can be derived from Eq.~\eqref{eq: H_secular_extra}, upon making in $\langle{\mathcal{H}}\rangle$ the substitution $\cos I=J_{\theta}/J_{\gamma}$. One has 
\begin{equation}\label{thetaequation}
    \frac{d\theta}{d\tau}=\frac{\partial \langle{\mathcal{H}}\rangle}{\partial J_{\theta}}=\Omega_{\rm g}-2\Omega_{\mbox{\tiny ZLK}}^{(\rm GR)}\cos I(1-e^2+5e^2\sin^2\gamma)~~,
\end{equation}
where it is evident the contribution of the gyroscope precession in the distant-star frame. Using the canonical transformation detailed in Sec.~\ref{sec:marck_variables} it is immediate to derive an analogous equation for the shifted angle $\theta'=\theta-\Omega_g \tau$ in Marck's frame, which indeed lacks the gyroscope precession contribution. We stress that in passing from the distant-star to Marck's frame of reference only $\theta$ changes in $\theta'$, and therefore all other equations of motion written before remain unaltered. 
\lbreak

Finally, the equation of motion for the argument of the periapsis $\gamma$ can be found by trading the eccentricity $e$ for the angular momentum $J_{\gamma}$ and using again  that $\cos I=J_{\theta}/J_{\gamma}$. We get
\begin{align}
\label{eq:gammadot}
    \frac{d\gamma}{d\tau}&=\frac{\partial \langle{\mathcal{H}}\rangle}{\partial J_{\gamma}}=2\Omega_{\mbox{\tiny ZLK}}^{(\rm GR)}[2(1-e^2)-5(1-e^2-\cos^2I)\sin^2\gamma]~~.
\end{align}

\subsection{General-relativistic effects in the ZLK mechanism}
\label{sec: rel_effects}

The ZLK mechanism is an effect for which a binary system under the influence of the tidal forces of an outer third body can exhibit a periodic exchange of eccentricity and orbital inclination, with a timescale much larger than its orbital period \cite{1962AJ.....67..591K,1976CeMec..13..471L,doi:10.1146/annurev-astro-081915-023315}.
\lbreak

To understand under which circumstances this mechanism can operate, one can start by recalling that, in the Newtonian approximation for the inner binary, the secularly-averaged Hamiltonian $\langle\mathcal{H}\rangle$ and the angular momentum projection $J_\theta$ are conserved quantities. Their values can therefore be estimated by fixing initial conditions for the eccentricity $e_0$ and the inclination $I_0$. The corresponding values for $\langle\mathcal{H}\rangle$ and $J_\theta$  will be labeled as $\langle\mathcal{H}\rangle_0$ and $({J_\theta})_0 $. By restricting to the case in which the inner orbit is initially circular, $e_0=0$, from the conservation of energy and angular momentum, explicitly $\langle\mathcal{H}\rangle_0=\langle\mathcal{H}\rangle_{\gamma=\pi/2}$ and $(J_\theta)_0=\mu \sqrt{G M a (1-e_{\rm max}^2)}\cos I_{\rm min}$, one gets
\begin{equation}\label{emax}
    e_{\rm max}=\sqrt{1-\frac{5}{3}\cos^2I_0}~~,~~\cos I_{\min}=\pm{\sqrt{\frac{3}{5}}}~~.
\end{equation}
Notice that, being $I_{\rm min}$ independent of the initial inclination $I_0$, it does not only constitute the minimum inclination reached in the ZLK oscillation but also the critical angle for the onset of the ZLK effect \cite{doi:10.1146/annurev-astro-081915-023315}.
For the ZLK mechanism to work, one has that the initial inclination $I_0$ should obey $\abs{\cos I_0}<\sqrt{3/5}$. For an initial inclination $I_0\approx\pi/2$ of the inner  binary system in circular orbit ($e_0=0$), the system  exhibits high eccentricity $e_{\rm max}\approx 1$. 

We thus find that including full general relativistic effects for the outer orbit does not alter the way the ZLK mechanism is triggered, it only affects the frequencies associated with the ZLK oscillations, as discussed in Sec.~\ref{sec: first_look}. In Sec.~\ref{sec: GW} we will include PN effects in the inner binary dynamics that will modify the condition \eqref{emax} \cite{PhysRevD.89.044043}.
\lbreak

A more exhaustive way to see how the ZLK mechanism manifests itself together with the general-relativistic gyroscope precession consists of considering the interchange between the eccentricity vector $\boldsymbol{e}$ of the inner binary and its angular momentum $\boldsymbol{L}_{\rm in}$ relative to the orbital plane of the outer binary. More specifically, from Eqs.~\eqref{eq: e_vec},\eqref{Lin} and \eqref{eq: J_def2}, one has 
\begin{equation}
    \frac{d\boldsymbol{L}_{\rm in}}{d\tau}=\frac{dL_{\rm in}}{d\tau}\boldsymbol{\hat L}_{\rm in}+L_{\rm in}\frac{d\boldsymbol{\hat L}_{\rm in}}{d\tau}~~~,~~~
    \frac{d\boldsymbol{e}}{d\tau}=\frac{de}{d\tau}\boldsymbol{\hat u}+e\frac{d\boldsymbol{\hat u}}{d\tau}~~.
\end{equation}
These evolution equations can be computed explicitly by recalling that both directions for $\boldsymbol{e}$ and $\boldsymbol{L}_{\rm in}$ are parametrized in terms of Euler angles $I$, $\gamma$, and $\theta$ (see Eqs.~\eqref{eq: uv},~\eqref{eq: e_vec}, and \eqref{Lin}) whereas their magnitudes are respectively related to the eccentricity $e$ and the Delauney variable $J_\gamma$ (see Eqs.~\eqref{eq: J_def2} and \eqref{dedt}) so that
\begin{align}
\label{eq: dLdt_dedt_1}
    \frac{d\boldsymbol{L}_{\rm in}}{d\tau}&=\frac{dJ_{\gamma}}{d\tau}\boldsymbol{\hat L}_{\rm in}+J_{\gamma}\left(\frac{d\boldsymbol{\hat L}_{\rm in}}{dI}~\frac{dI}{d\tau}+\frac{d\boldsymbol{\hat L}_{\rm in}}{d\theta}~\frac{d\theta}{d\tau}\right)~~,
    \\
\label{eq: dLdt_dedt_2}
    \frac{d\boldsymbol{e}}{d\tau}&=\frac{de}{d\tau}\boldsymbol{\hat u}+e\left(\frac{d\boldsymbol{\hat u}}{dI}~\frac{dI}{d\tau}+\frac{d\boldsymbol{\hat u}}{d\theta}~\frac{d\theta}{d\tau}+\frac{d\boldsymbol{\hat u}}{d\gamma}~\frac{d\gamma}{d\tau}\right)~~.   
\end{align}
By exploiting vector identities,  
Eqs.~\eqref{eq: dLdt_dedt_1} and \eqref{eq: dLdt_dedt_2} can be rewritten as
\begin{align}
    \nonumber
    \frac{d\boldsymbol{L}_{\rm in}}{d\tau}=& ~\boldsymbol{ \Omega}_{g}\times\boldsymbol{ L}_{\rm in}+2\Omega_{\mbox{\tiny ZLK}}^{(\rm GR)} J_\gamma\Big[(1-e^2)(\boldsymbol{ \hat L}_{\rm in}\cdot\boldsymbol{\hat L}_{\rm out})(\boldsymbol{ \hat L}_{\rm in}\times\boldsymbol{\hat L}_{\rm out})
    \\ \label{Linequation}
    &-5(\boldsymbol{ e}\cdot\boldsymbol{\hat L}_{\rm out})(\boldsymbol{ e}\times\boldsymbol{\hat L}_{\rm out})\Big]~~,
\\ \nonumber 
    \frac{d\boldsymbol{e}}{d\tau}=&~\boldsymbol{\Omega}_{g}\times \boldsymbol{e}+
   2\Omega_{\mbox{\tiny ZLK}}^{(GR)}\Big[(\boldsymbol{\hat L}_{\rm in}\cdot \boldsymbol{\hat L}_{\rm in})(\boldsymbol{e}\times\boldsymbol{\hat L}_{\rm in})+2 \boldsymbol{\hat L}_{\rm in}\times \boldsymbol{e}
   \\ \label{eequation}
   &-5(\boldsymbol{e}\cdot \boldsymbol{\hat L}_{\rm out})(\boldsymbol{\hat L}_{\rm in}\times\boldsymbol{\hat L}_{\rm out})\Big]~~,
\end{align}
where we used $\boldsymbol{\Omega_g}=\Omega_g \boldsymbol{\hat L}_{\rm out}$ and the results obtained in Eqs.~\eqref{dedt} and \eqref{eq: Jgamma1dot}.
\lbreak

Eqs.~\eqref{Linequation} and \eqref{eequation} are written in terms of the proper time $\tau$ related to the inner binary system.
From the point of view of an asymptotic observer, they become
\begin{align} \nonumber
    \frac{d\boldsymbol{L}_{\rm in}}{d \hat{t}}=& ~\boldsymbol{ \Omega}_{g}^{(\infty)}\times\boldsymbol{ L}_{\rm in}+2\Omega_{\mbox{\tiny ZLK}}^{(\infty)} J_\gamma\Big[(1-e^2)(\boldsymbol{ \hat L}_{\rm in}\cdot\boldsymbol{\hat L}_{\rm out})(\boldsymbol{ \hat L}_{\rm in}\times\boldsymbol{\hat L}_{\rm out})
    \\\label{Linasequation}
    &-5(\boldsymbol{ e}\cdot\boldsymbol{\hat L}_{\rm out})(\boldsymbol{ e}\times\boldsymbol{\hat L}_{\rm out})\Big]~~,
    \\ \nonumber
    \frac{d\boldsymbol{e}}{d \hat{t}}=&~\boldsymbol{\Omega}_{g}^{(\infty)}\times \boldsymbol{e}+
   2\Omega_{\mbox{\tiny ZLK}}^{(\infty)}\bigg[(\boldsymbol{\hat L}_{\rm in}\cdot \boldsymbol{\hat L}_{\rm in})(\boldsymbol{e}\times\boldsymbol{\hat L}_{\rm in})+2 \boldsymbol{\hat L}_{\rm in}\times \boldsymbol{e}
   \\  \label{easequation}
   &-5(\boldsymbol{e}\cdot \boldsymbol{\hat L}_{\rm out})(\boldsymbol{\hat L}_{\rm in}\times\boldsymbol{\hat L}_{\rm out})\bigg]~~,
\end{align}
where we defined
\begin{equation}
    \Omega_g^{(\infty)}=\frac{1}{u^t}\Omega_g~.
\end{equation}
Notice that the redshifted gyroscope precession frequency above is always finite: at the ISCO, for instance, one has $\Omega_g^{(\infty)}=(c/\hat r_{\rm ISCO}) (\sqrt{2}-1)/(2\sqrt{3})$ in the non-spinning case $\chi=0$, whereas at extremality, $\chi=1$, one finds $\Omega_g^{(\infty)}=1/2 (c/r^+_{\rm ISCO})$ and $\Omega_g=(c/\hat r^-_{\rm ISCO})(4/\sqrt{3}-9)/26$ respectively for the co-rotating and counter-rotating orbits.

Eq.s~\eqref{Linasequation} and \eqref{easequation} include the gyroscope precession and extend the results previously known in the literature for the ZLK effect \cite{2015MNRAS.447..747L,Liu_2019, Kuntz:2022onu} to the case in which the external body, in this case a supermassive Kerr black hole, is described using the Kerr metric \eqref{KerrBL}, thus being in a strong GR regime. 
\lbreak

In the case of a circular orbit for the inner binary, $\boldsymbol{e}=0$, which constitutes a solution for the equation for the eccentricity, Eq.~\eqref{Linasequation} becomes 
\begin{equation}
\label{Linasequationcir}
    \frac{d\boldsymbol{L}_{\rm in}}{d\hat{t}}=\left[\Omega_{g}^{(\infty)}-2\Omega_{\mbox{\tiny ZLK}}^{(\infty)}(\boldsymbol{ \hat L}_{\rm in}\cdot\boldsymbol{\hat L}_{\rm out})\right](\boldsymbol{ \hat L}_{\rm out}\times\boldsymbol{ L}_{\rm in})~~,
\end{equation}
which describes the precession of the angular momentum of the inner binary $\boldsymbol{L}_{\rm in}$ around the direction of the angular momentum of the outer binary $\boldsymbol{\hat L}_{\rm out}$. 
In the weak-field limit $\hat r\to \infty$, one has
\begin{equation}
\begin{split}
 &\frac{d\boldsymbol{L}_{\rm in}}{d\hat{t}}=
 \\
 &\bigg[ \frac{3}{2}\frac{(G m_3)^{\frac{3}{2}} \sigma   }{c^2~\hat{r}^{\frac{5}{2}}}-\frac{G J_3 }{c^2 \hat r^3}-\frac{3}{4}\frac{m_3}{\hat r^3}\sqrt{\frac{G a^3}{M}}(\boldsymbol{\hat L}_{\rm in}\cdot\boldsymbol{\hat L}_{\rm out})\bigg](\boldsymbol{\hat L}_{\rm out}\times\boldsymbol{ L}_{\rm in})      
 \\\nonumber
 &+
 \bigg[ \frac{9}{8}\frac{(G m_3)^{\frac{5}{2}} \sigma   }{c^4~\hat{r}^{\frac{7}{2}}}-3\frac{G^2 m_3 J_3}{c^4 \hat r^4}-\frac{9}{8}\frac{G m_3^2}{c^2 \hat r^4}\sqrt{\frac{G a^3}{M}}(\boldsymbol{\hat L}_{\rm in}\cdot\boldsymbol{\hat L}_{\rm out})\bigg](\boldsymbol{\hat L}_{\rm out}\times\boldsymbol{ L}_{\rm in})
\\\nonumber
 &+\mathcal{O}  \left(\hat r^{-9/2}\right)~~,
\end{split}
\end{equation}
where in the first line, the first term represents the 1PN contribution due to the gyroscope precession, the second term is a relativistic effect related to the spin of the SMBH and the third term represents the precession generated by the standard ZLK mechanism. This is consistent with the results presented in Ref.~\cite{Liu_2019}, in the hierarchical regime for circular outer orbits and where the contribution proportional to the spin of the SMBH comes from the Lense-Thirring precession.
The second line is instead a new result and represents higher-order contributions which we predict using the result in Eq.~\eqref{Linasequationcir}.

\begin{figure*}
    \centering
    \includegraphics[width=0.9\textwidth]{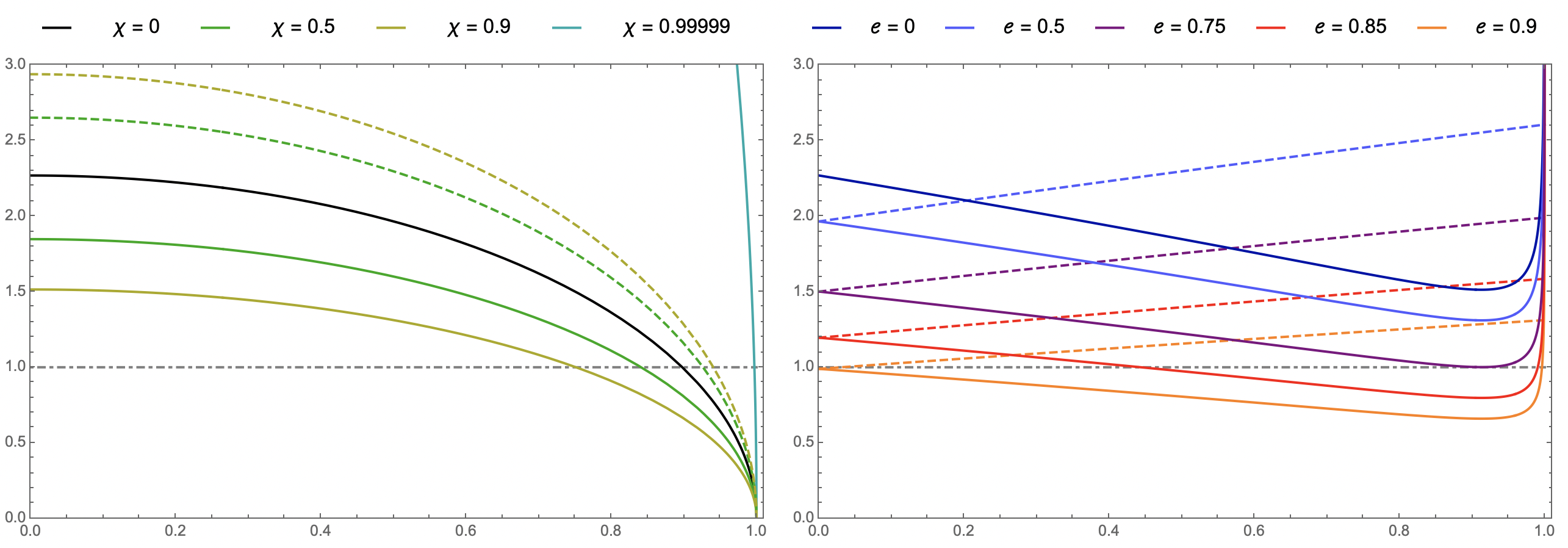}
    \begin{picture}(0,0)
        \put(-115,-5){$\chi$ }
        \put(-345,-5){$e$ }
        \put(-485,75)
        {{{$\bigg|\dfrac{\Omega_{\mbox{\tiny g}}}{\Omega_{\mbox{\tiny ZLK}}}\bigg|$}}}
    \end{picture}
    \caption{In this figure we set $m_1=m_2=10~ M_\odot$, $m_3=2\times10^9~M_\odot$ and $a=0.1 ~\rm{AU}$. (Left panel) 
    Absolute value of the ratio between the gyroscope precession frequency and the ZLK frequency as a function of the inner binary eccentricity $e$. The different curves are obtained for different values of the dimensionless spin parameter $\chi$ of the SMBH. (Right panel) Absolute value of the ratio between the gyroscope precession frequency and the ZLK frequency as a function of the dimensionless spin parameter $\chi$ of the SMBH, for different values of the eccentricity of the inner binary $e$. The solid and dashed coloured curves are respectively for co-rotating ($\sigma=+1$) and counter-rotating ($\sigma=-1$) outer geodesics. The value $R=1$, depicted with the dot-dashed line, marks the trans-adiabatic regime in which $\Omega_g^{(\infty)}\approx \Omega_{\rm ZLK}^{(\infty)}$. The rightmost curve in the left panel, corresponding to $\chi=0.99999$, is reported to show that $R\to \infty$ for $\chi\to 1$, in agreement with the behaviour presented in the right panel and the discussion made below Eq.~\eqref{eq: Omegag}. }
    \label{fig: Plref}
\end{figure*}
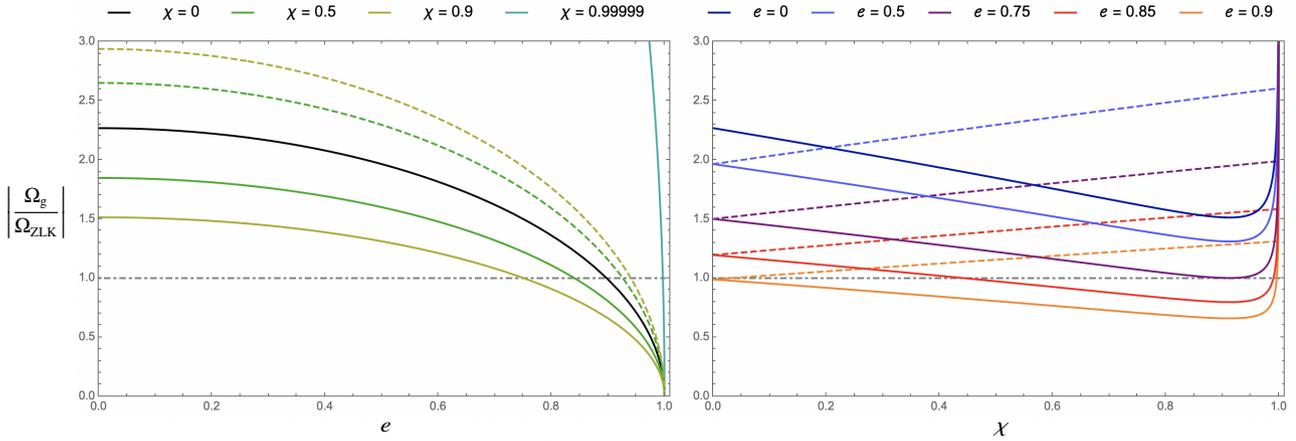

As pointed out in the past literature (see for instance \cite{Petrovich:2017otm}, \cite{Liu:2017yhr} and \cite{Liu_2019} ) the interplay of the ZLK mechanism with additional precessing effects can lead to significant alteration in the BBH dynamics and exhibit chaotic features. The dynamical behaviour can be identified by means of an adiabatic parameter, that we define as 
\begin{equation}
    R\equiv\abs{\frac{\Omega_{g}^{(\infty)}}{\Omega_{\mbox{\tiny ZLK}}^{(\infty)}}}\Bigg\vert_{\hat r^\sigma_{\rm ISCO}}=\abs{\frac{\Omega_g}{\Omega_{\mbox{\tiny ZLK}}^{(\rm GR)}}}\Bigg\vert_{\hat r^\sigma_{\rm ISCO}}~~,
\end{equation}
where the ratio is evaluated at the ISCO, since $\hat r^\sigma_{\rm ISCO}$ marks the scale at which the strong-gravity effects are more relevant in our setup. %
Fig.~\ref{fig: Plref} shows plots for the ratio $R$ in a specific configuration for the BBH-SMBH system as a function of the inner BBH system eccentricity $e$ and the SMBH spin parameter $\chi$. 
It would be interesting to study the interplay between the gyroscope precession and the ZLK mechanism further.

\section{Binary merger time close to a supermassive BH}
\label{sec: GW}

In this section, we refine the dynamics of the inner binary system by adding the periastron precession and GW emission, so that we can study the interplay of these effects together with the ZLK mechanism. This enables us to study how treating the SMBH in strong gravity can alter the dynamics of the BBH system compared to if one included only the Newtonian gravity effect of the SMBH~\cite{Blaes:2002cs,Randall:2018qna}.
\lbreak

We recall that to treat the presence of the SMBH as a perturbation of the BBH system the condition \eqref{SEPtidal} must be satisfied. After introducing the parametrization for an elliptic orbit as in \eqref{eq: r_ea}, the tidal condition can be rewritten in terms of the semi-major axis $a$ of the binary system as 
\begin{equation}
\label{eq:tidal_cond}
    a \ll \hat{r} \sqrt{
\frac{c^2 \hat{r}}{Gm_3}} \,.
\end{equation}
This ensures that we can safely neglect the GW
backreaction of the outer orbit and consider therefore only the GWs emitted by the inner BBH~\cite{Randall:2018qna}.

\subsection{Post-Newtonian dynamics of the binary system}

GWs emitted by the inner BBH system reduce its energy and angular momentum and consequently its semi-major axis $a$ and eccentricity $e$. Peters's equations \cite{PhysRev.131.435,1964PhRv..136.1224P} keep into account this gravitational backreaction by providing the orbit-averaged evolution of  $e$ and  $a$ for an isolated binary. The average variations read
\begin{equation}
\label{eq: PetEQ}
    \begin{split}
        \left\langle\frac{d a}{d \tau}\right\rangle_{\rm GW}=&-\frac{64}{5}\frac{  G^3 \mu  M^2}{  a^3 c^5}\frac{1}{\left(1-e^2\right)^{7/2}}\left(1+\frac{73 }{24}e^2+\frac{37 }{96}e^4\right)~~,
        \\
        \left\langle\frac{d e}{d \tau}\right\rangle_{\rm GW}=&-\frac{304}{15}\frac{  G^3 \mu  M^2}{  a^4 c^5}\frac{e}{\left(1-e^2\right)^{5/2}}\left(1+\frac{121 }{304}e^2\right)~~.
    \end{split}
\end{equation}
These equations were obtained to model the gravitational backreaction for an isolated binary, but one can include the influence of a third external body, that interacts with the binary through the ZLK mechanism, by adding these PN contributions to the system of equations describing the time evolution of the orbital variables $(a,e,\gamma,I)$, as given in Eqs.~ \eqref{eq: I_evolution}, \eqref{eq: e_evolution}, \eqref{eq:gammadot}. 
In particular, the first of the two PN contributions above is responsible for triggering the inspiral phase for the binary system, by decreasing the relative distance $a$ between the two masses. As opposed to the ZLK mechanism, which can lead to an increase of the orbital eccentricity $e$, the second equation shows how the emission of GWs tends to circularise the orbit. 
\lbreak

We recall that $J_\gamma$ represents the magnitude of the angular momentum for a Newtonian binary system. Hence, by combining Eqs.~\eqref{eq: PetEQ} with the definition of $J_\gamma$, according to Eq.~\eqref{eq: J_def1}, it is immediate to estimate the loss of angular momentum associated to the GW emission, namely~\cite{Landau:1975pou}
\begin{equation}
\label{eq: dJdt_GW}
    \left\langle\frac{d J_\gamma}{d \tau}\right\rangle_{\rm GW}=-\frac{32}{5}\frac{G^{7/2} \mu ^2 M^{5/2}}{ a^{7/2} c^5}\frac{1}{(1-e^2)^2}\left(1+\frac{7 }{8}e^2\right)~~.
\end{equation}
It is important to stress that the gravitational backreaction only reduces the magnitude of the angular momentum, without affecting its direction. 
In other words, if we ignore the presence of the SMBH, the binary system remains in the same plane when including the GW emission. 
Thus, we have $\langle dI/d\tau\rangle_{\rm GW}=0$ which from Eq.~\eqref{eq: dIdt} gives
\begin{equation}
    \left\langle\frac{d J_\theta}{d \tau}\right\rangle_{\rm GW}=\left\langle\frac{d J_\gamma}{d \tau}\right\rangle_{\rm GW}\cos I~~.
\end{equation}

It is immediate to notice, using Eq.~\eqref{eq: dJdt_GW}, that the loss of angular momentum (or, analogously, the decreasing of the orbital distance and eccentricity, according to Eq.~\eqref{eq: PetEQ}) due to GW emission becomes extremely efficient when the condition $e\approx 1$ is met. As we discuss this in more detail in the following subsection, the enhancement in the eccentricity due to the ZLK mechanism can therefore boost the merger of eccentric binary systems. While such \textit{ZLK-boosted mergers} were already noticed in previous works that analyzed the combined ZLK dynamics with PN effects (see for instance Refs.~\cite{Blaes:2002cs,Randall:2018qna}), in the next subsection we show that strong gravity effects associated to the external SMBH can lead to further significant changes in the frequency of the ZLK-oscillations and in the merger time.
\lbreak

Finally, following \cite{Randall:2018qna}, we include also the effect of the periastron precession of elliptic orbits as an additional contribution to the evolution equations, since this 1PN effect plays a significant role in the ZLK mechanism. Indeed, as we shall review later, it is known to limit the range in which the ZLK mechanism is valid. Notice that the periastron precession preserves the angular momentum vector $d\boldsymbol{L}_{\rm in}/d\tau=0$, meaning for instance that the orbital plane remains unchanged and only manifests itself as an apsidal advance $d\boldsymbol{\hat e}_{\rm in}/d\tau=(d\gamma/d\tau)\boldsymbol{\hat v}$ (see Eqs.~\eqref{eq: dLdt_dedt_1} and \eqref{eq: dLdt_dedt_2}). The periastron precession can therefore be written in terms of a first derivative for the argument of periapsis $\gamma$, according to 
\begin{equation}
\label{eq: PNP}
    \left(\frac{d \gamma}{d \tau}\right)_{PN}=\frac{3 }{a c^2 \left(1-e^2\right)}\left(\frac{G M}{a}\right)^{3/2}~~.
\end{equation}
As discussed below, these PN effects alter the dynamics of the ZLK mechanism for BBH systems influenced by the presence of an external mass.
\lbreak

We can now write down a closed set of equations by combining the evolution equations Eqs.~\eqref{eq: I_evolution}, \eqref{eq: e_evolution}, \eqref{eq:gammadot} of Sec.~\ref{sec: evo_eqs}, which considers the strong gravity effects of an external spinning SMBH, with the effects of GW emission and periastron precession introduced above. 
This gives the evolution equations
\begin{align}
\label{eq:PNGWeqs}
\begin{split}
    \left\langle\frac{d a}{d \tau}\right\rangle=&-\frac{64}{5}\frac{  G^3 \mu  M^2}{  a^3 c^5}\frac{1}{\left(1-e^2\right)^{7/2}}\left(1+\frac{73 }{24}e^2+\frac{37 }{96}e^4\right)~~,
    \\
    \left\langle\frac{d e}{d \tau}\right\rangle=&~5\Omega_{\mbox{\tiny ZLK}}^{(\rm GR)}e(1-e^2)\sin^2I \sin2\gamma
    \\
    &-\frac{304}{15}\frac{  G^3 \mu  M^2}{  a^4 c^5}\frac{e}{\left(1-e^2\right)^{5/2}}\left(1+\frac{121 }{304}e^2\right)~~,
    \\
    \left\langle\frac{d \gamma}{d \tau}\right\rangle=&~2\Omega_{\mbox{\tiny ZLK}}^{(\rm GR)}[2(1-e^2)-5(1-e^2-\cos^2I)\sin^2\gamma]
    \\
    &+\frac{3 }{a c^2 \left(1-e^2\right)}\left(\frac{G M}{a}\right)^{3/2}~~,
    \\
    \left\langle\frac{d I}{d\tau}\right\rangle=&-\frac{5}{2}\Omega_{\mbox{\tiny ZLK}}^{(\rm GR)} e^2\sin 2I \sin2\gamma~~,
\end{split}
\end{align}
where the equations have to be supplemented with the definition $J_\gamma=\mu\sqrt{G M a (1-e^2)}$, as in Eq.~\eqref{eq: J_def1}, and with the general-relativistic definition of the ZLK frequency $\Omega_{\mbox{\tiny ZLK}}^{(\rm GR)}$, that we derived in Eq.~\eqref{eq: Omega_KL}. 
These evolution equations are written with respect to the proper time $\tau$, being the time associated with the inner binary system. However, one can use $\tfrac{d}{d\hat{t}}= \tfrac{1}{u^t} \tfrac{d}{d\tau}$ to translate these equations into evolution equations with respect to the asymptotic time $\hat{t}$ to obtain the dynamical description as seen by an asymptotic observer.
\lbreak

The system of equations \eqref{eq:PNGWeqs} allows us to study how the orbital parameters $(e,a,I,\gamma)$ of the inner binary,  evolve in time. To solve them, one has to specify initial conditions for the inner BBH system, $(a_0,e_0,\gamma_0,I_0)$, a set of parameters that characterize the outer circular equatorial orbit around the Kerr SMBH $(\hat r,s_3,\sigma=\pm1)$, and the three masses $(m_1,m_2,m_3)$.
\lbreak

In the next subsection, we solve numerically the system of equations \eqref{eq:PNGWeqs}, for some specific case of interest.
We will work in astronomical units, i.e. (AU, $M_{\odot}$ and years), where $G=4 \pi^2$ and $c=63072\hspace{1mm} \text{AU}/\text{years}$.
\lbreak

As previously explained, we need to impose the condition~\eqref{eq:tidal_cond} for the tidal approximation to be valid. 
This allows us to neglect the GW backreaction of the outer orbit over the timescale that governs the inner binary dynamics.
Furthermore, we are working in the near-Newtonian regime \eqref{SEPbinary} in which we can treat the black holes in the inner binary approximately as particles. 
\lbreak

The ZLK oscillations take place on a timescale much longer than the orbital period of the binary system around the SMBH. Thus, it becomes important to ensure that the triple system is stable, allowing the ZLK mechanism to enhance the eccentricity of the BBH system. When the binary system is too close to the SMBH, the interaction between the three bodies can result in a tidal breakup. 

\begin{figure}
    \centering
    \includegraphics[width=0.4\textwidth]{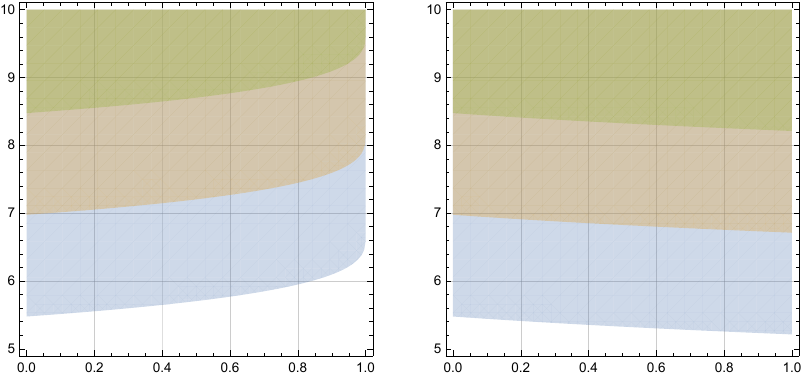}
    \begin{picture}(0,0)
        \put(-160,-3){\tiny{$\chi$} }
        \put(-50,-3){\tiny{$\chi$} }
        \put(-215,30){\tiny{\rotatebox{90}{{$\log_{10}(m_3/2 M_{\odot})$}}}}
        \put(-105,30){\tiny{\rotatebox{90}{{$\log_{10}(m_3/2 M_{\odot})$}}}}
    \end{picture}
    \caption{The colored regions represent regions of the parameter space for the SMBH which are consistent with the tidal breakup condition $\hat r/a>(3m_3/M)^{1/3}$. The green, orange and blue colors respectively identify $\hat r=\hat r^\sigma_{\rm ISCO}$, $\hat r=10~\hat r^\sigma_{\rm ISCO}$ and $\hat r=100~\hat r^\sigma_{\rm ISCO}$, whereas the left and right panels distinguish co-rotating ($\sigma=+1$) and counter-rotating orbits  ($\sigma=-1$). The plots are obtained by fixing $m_1=m_2=10~M_\odot$ and $a=0.1~\rm{AU}$, as representative values for a stellar-mass BBH system.}  
    \label{fig:enter-label}
\end{figure}
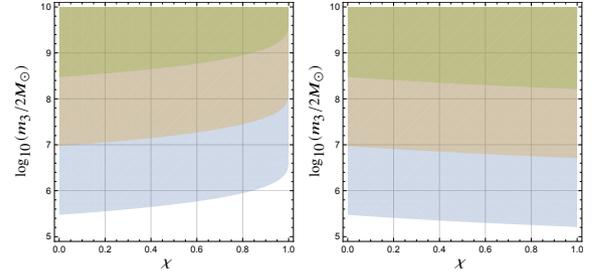

To ensure that the presence of an SMBH does not lead to a tidal break up of the binary system, the following condition must be satisfied~\cite{Miller_2005,Antonini_2012}
\begin{equation}
    \label{eq:breakup}
    \hat{r} \geq \hat r_{\rm{tide}} \approx a \left(3 m_3/M\right)^{1/3},
\end{equation}
where we recall that $M=m_1+m_2$ is the total mass of the BBH system and $m_3$ the mass of the SMBH perturbing the binary.
In our analysis, we carefully chose the radial distance between the binary system and the SMBH so as to always guarantee the stability of the triple system, more specifically in all our examples, we will consider $\hat r \gg \hat r_{\rm{tide}}$. As an illustrative example, in  Fig.~\ref{fig:enter-label} we plot the regions of the parameter space for the SMBH which are consistent with the tidal breakup condition \eqref{eq:breakup}, in the case of a stellar-mass BBH system with masses $m_1=m_2=10~M_\odot$ and separation $a=0.1~\rm{AU}$.

To prepare for the analysis we perform in the remaining of this section, we now briefly introduce respectively the GW and PN timescales at play in our investigation and the peak frequency of GW emitted by the BBH system. These will play a key role in the next subsections where we show how the GR effects in the strong field regime affect the merger time of the binary system and the emission of GW waves. 

\subsubsection{Time Scales}
For an isolated BBH system the merger time due to the emission of GWs is~\cite{1964PhRv..136.1224P,Maggiore:2007ulw}
\begin{equation}
    T_{\rm GW}=\frac{5}{256}\frac{c^5 a_0^4}{G^3 m_1 m_2 M}G(e_0)(1-e_0^2)^{7/2}~~, 
\end{equation}
where 
\begin{equation}
\begin{aligned}
    G(e_0)&=\frac{48}{19 g^4(e_0)(1-e_0^2)^{7/2}}\int_0^{e_0} \frac{g^4(\tilde e)(1-\tilde e^2)^{5/2}}{\tilde e\left(1+\frac{121}{304}\tilde e^2\right)}d\tilde e~~,
    \\
    g(e)&=\frac{e^{12/19}}{1-e^2}\left(1+\frac{121}{304}e^2\right)^{870/2299}~~.
\end{aligned}
\end{equation}
The function $G(e_0)\in [0.979,1.81]$ for $e_0\in[0,1]$, and for an order of magnitude estimate it can be ignored. For all practical purposes, a good approximation for an isolated BBH system merger time is thus given by 
\begin{align}
\label{eq: TGW}
    T_{\rm GW}&\approx 1.6~(1-e_0^2)^{7/2}\times 10^{10}~{\rm yrs}
    \\\nonumber
    &\quad\times\left(\frac{10 M_\odot}{m_1}\right)\left(\frac{10 M_\odot}{m_2}\right)\left(\frac{20 M_\odot}{M}\right)\left(\frac{a_0}{0.1~{\rm AU}}\right)^4~~.
\end{align}
The prefactor $(1-e_0^2)^{7/2}$ shows that for very high values of the eccentricity, the merger time can be drastically reduced. Tidal interactions generated by an external SMBH, through the ZLK mechanism, can produce high eccentricities, thus  catalyzing the coalescence for highly-inclined BBHs and speeding up the merger to timescales much shorter than those characteristic of isolated binaries with the same masses and relative distance \cite{Antonini:2012ad}.
\lbreak

From Eq.\eqref{eq: PNP} we can compute also the time scale associated with the periastron precession as
\begin{equation}
  T_{\rm PN}=  \frac{a c^2 \left(1-e^2\right)}{3 }\left(\frac{a}{G M}\right)^{3/2}\equiv\frac{2\pi}{\Omega_{\rm PN}}~~.
\end{equation}
In this regard, it is interesting to notice that, when the periastron timescale becomes comparable with the ZLK timescale $T_{\mbox{\tiny ZLK}}=2\pi/\Omega_{\mbox{\tiny ZLK}}$, the periastron precession destroys the ZLK resonance and the binary system begins to evolve as if it was isolated. To see the competing effects of the periastron precession and the ZLK mechanism, it is convenient to neglect the GW backreaction and analyze under which conditions $\langle d\gamma/ d\tau\rangle=0$ and $\langle d e/d\tau\rangle=0$ in Eq. \eqref{eq:PNGWeqs}, where the second equation is automatically satisfied by setting $\gamma=\pi/2$.  One finds  
\begin{equation}
\label{eq: KLPP}
    \cos^2 I_0=\frac{3}{5}(1-e^2)-\frac{3}{10}\frac{1}{(1-e^2)}\frac{\left(\frac{G M}{a}\right)^{3/2}}{a c^2 \Omega_{\mbox{\tiny ZLK}}^{(\rm GR)}}
\end{equation}
which replaces Eq.\eqref{emax} when the periastron precession is taken into account. Due to the presence of $\Omega_{\mbox{\tiny ZLK}}^{(\rm GR)}$, this generalizes the Newtonian formula found in  \cite{Blaes:2002cs}.
The window of values for the critical inclination that triggers the ZLK resonance is reduced by the presence of the periastron effect.
This makes it complicated to find a condition on $\hat{r}$ that would result in the condition $\cos^2 I_0>0$. 
To illustrate this, consider the simpler case of a non-spinning SMBH for which we find the following condition 
\begin{equation}
    \frac{\hat r^3}{a^3}\left(1-\frac{3 G m_3}{c^2 \hat r}\right)<\frac{3}{4}(1-e^2)^{3/2}\left(\frac{a c^2}{G M}\right)\frac{m_3}{M}~~~.
\end{equation}
In the limit in which this bound is saturated, the critical inclination value is close to $I_0\approx90^\circ$, so that the minimum and maximum values for the inclination almost coincide. It is clear from \eqref{eq: KLPP} that, consistently, the maximum value of the eccentricity is lowered compared to the case in which the periastron precession is neglected.

\subsubsection{Peak Frequency}

For eccentric binaries, the GW spectrum is spread across an infinite number of harmonics \cite{PhysRev.131.435}, with frequencies that are integer multiples $n$ of the fundamental Keplerian frequency $1/(2\pi) \sqrt{G M/a^3 }$, and peaked approximately at \cite{Wen:2002km} 
\begin{equation}
\label{eq: PeakGW}
    f_{\rm GW}\simeq\frac{\sqrt{G M}}{\pi[a (1-e^2)]^{3/2}}(1+e)^{1.1954}~~.
\end{equation}
In the evolution of an isolated binary the gravitational backreaction, encoded in Eqs.~\eqref{eq: PetEQ}, would contribute to circularize the orbit and move the peak frequency towards the usual $n=2$ harmonic long before the merger takes place \cite{Maggiore:2007ulw}.  
Under the influence of an external body, however, the ZLK mechanism can provide large eccentricity oscillations in highly-inclined binaries and can enhance the peak frequency to values high enough to enter in the sensitivity band of future space-based GW detectors \cite{Antonini_2012}.

\subsection{Binary merger time in the weak field limit}
\label{sec: GW_weak}

In Sec.~\ref{sec:gyroscope} we studied how GR effects induced by the presence of a spinning SMBH can lead to a significant enhancement of the frequency for the ZLK resonance in the case of a Newtonian binary system moving on a circular geodesic. By superimposing the ZLK mechanism with the PN dynamics of the inner BBH system, according to the discussion made in the previous subsection, we are now in a position to study how the BBH merger time is influenced by the presence of an external SMBH.
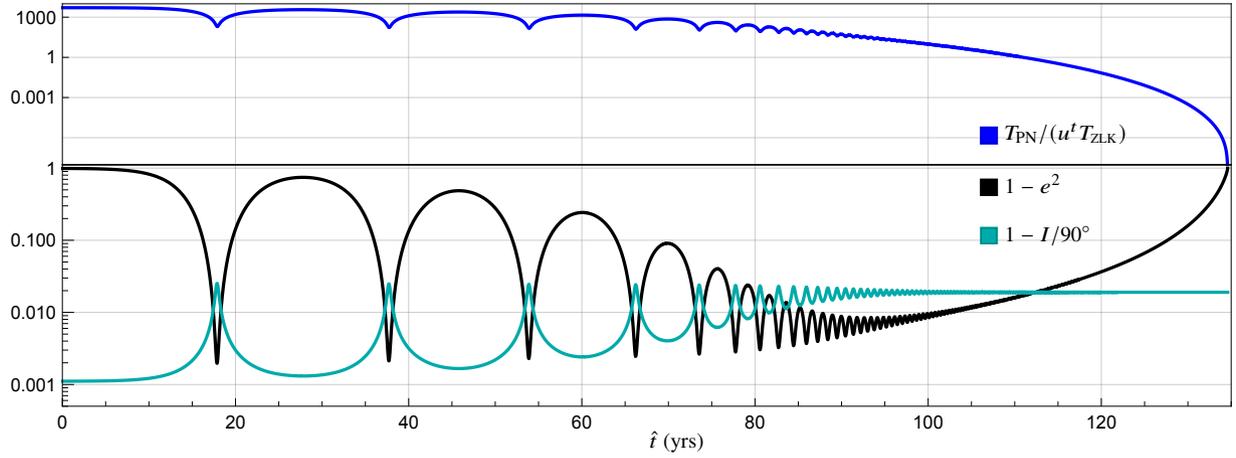
\begin{figure*}
    \centering
    \includegraphics[width=0.9\textwidth]{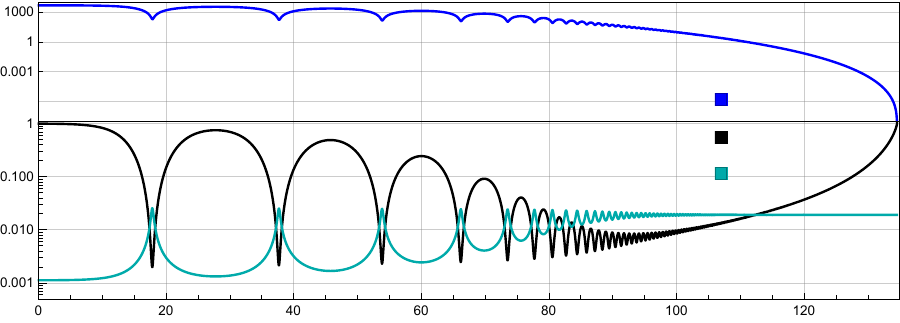}
    \begin{picture}(0,0)
        \put(-220,-3){$\hat t$ (yrs)}
        \put(-87,112){{\footnotesize{$T_{\rm PN}/(u^tT_{\mbox{\tiny ZLK}})$}}}
        \put(-87,92){{\footnotesize{$1-e^2$}}}
        \put(-87,74){{\footnotesize{$1-I/90^\circ$ }}}
    \end{picture}
    \caption{We focus on the spinless case for the SMBH, i.e. $\chi=0$. Here $\hat r =120 ~\rm{AU}$, which corresponds to approximately $\hat r\approx 500~\hat r_{\rm ISCO}$ for a non-spinning black hole with mass $m_3=4\times 10^{6}M_\odot$. We combine the effect of the ZLK mechanism with the periastron precession and GW emission. The picture highlights that the maximum values of the BBH eccentricity $e$ corresponds to the minimum values for its orbital inclination $I$ and viceversa.}
    \label{fig: KLspin}
\end{figure*}  

We will start the study of the evolution equations \eqref{eq:PNGWeqs} by considering a case in which we are in the weak gravity regime, regarding the influence of the SMBH on the binary system. 
This regime has already been studied previously, {\sl e.g.}~in \cite{Randall:2018qna}, but here we use it to provide a baseline for Section \ref{sec: GW strong}, where we find novel results for the strong field regime, also described by Eqs.~\eqref{eq:PNGWeqs}. 
\lbreak

We investigate the evolution equations \eqref{eq:PNGWeqs} by solving them numerically.
For the weak-field example of this section we choose fixed masses $m_1=m_2=10~ M_{\odot}$ and $m_3=4\times 10^6 ~M_{\odot}$ for the three black holes.%
\footnote{Note that the system has the scaling property that if we scale all three masses $m_1,m_2$, and $m_3$ with the same factor, along with the initial conditions for the semi-major axis of the inner binary $a_0$ and radius of the outer orbit $\hat r$, the merger time will be rescaled with this factor as well.}
We use these values to connect to previous literature ({\sl e.g.}~Ref.~\cite{Randall:2018qna}) and since $m_3$ is roughly the mass of Sagittarius A*.
Moreover, we choose the initial conditions for the inner binary to be $e_0=0.1$, $a_0=0.1~\rm{AU}$ , $\gamma_0=0^\circ$ and $I_0=89.9^\circ$, so that we only need to specify the outer orbit parameters.%
\footnote{The initial value for the inclination angle $I$ is chosen specifically to maximize the ZLK mechanism. The initial value for the semi-major axis $a$ is fixed to be $0.1 \hspace{1mm} \text{AU}$. This value is small enough to ensure that the binary system does not break up when interacting with the SMBH, according to Eq.~\eqref{eq:breakup}, but it is also sufficiently large to allow the two black holes in the inner binary to evolve without merging too fast \cite{Miller_2005}.} 
\lbreak

For a non-spinning SMBH with the above-mentioned choice for $m_3$ we have the ISCO radius $\hat r_{\rm ISCO}=6~ G m_3/c^2\approx 0.24~\rm{AU}$. 
Naturally, the strong gravity effects from the presence of the SMBH are most significant at the ISCO.
We choose here the outer orbit radius to be $\hat r = 120~\rm{AU} \approx 500~\hat r_{\rm ISCO}$ which means we are within the weak-field regime (concerning the gravitational influence of the SMBH on the binary system).
In this regime, the redshift factor is $u^t\approx 1$ according to Eq.~\eqref{eq: redshift}.
At the same time, $\hat r = 120~\rm{AU}$ ensures we are well below the tidal breakup limit \eqref{eq:breakup}.
\lbreak

From Eq.~\eqref{eq: TGW} one can observe that, with our choice for the orbital parameters, it would take approximately $\sim 10^{10}$ years for an isolated binary  to coalesce. 
In Fig.~\ref{fig: KLspin} we show our numerical solution, which presents the characteristic quasi-periodic behavior of the BBH system influenced by an external SMBH. From Fig.~\ref{fig: KLspin} it is immediate to see that the time for the merger to occur is drastically reduced to order  $\sim 10^2$ years  by the presence of the eccentricity oscillations associated with the weak-gravity ZLK mechanism. 
\lbreak

\begin{figure}
    \centering
    \includegraphics[width=0.4\textwidth]{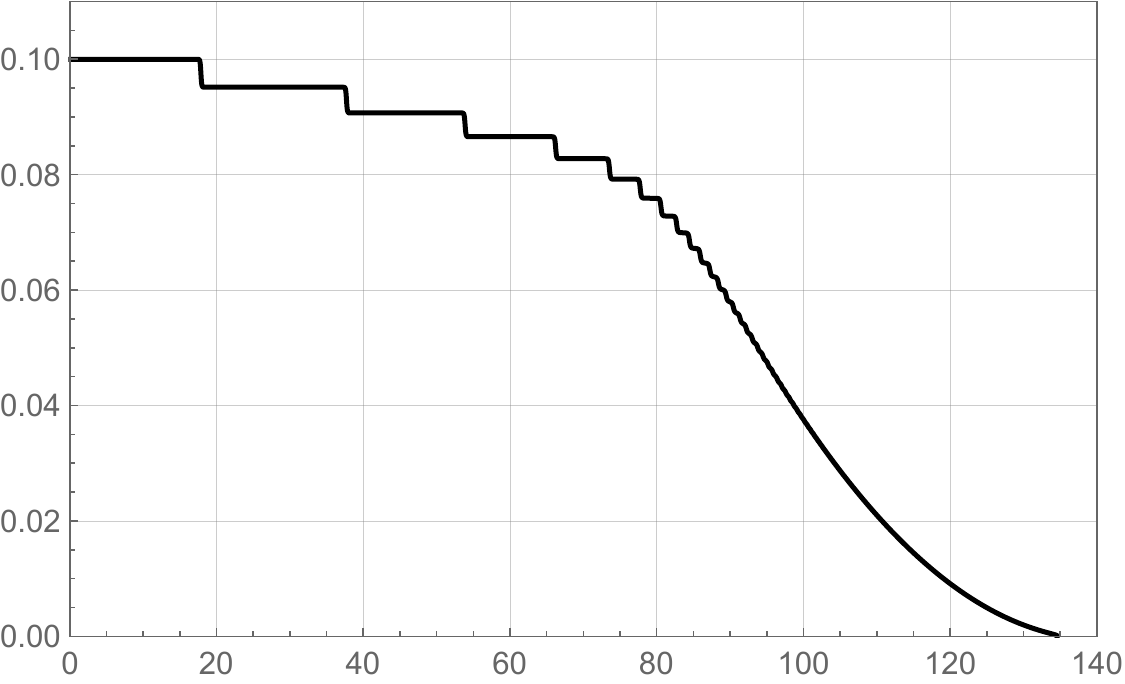}
    \begin{picture}(0,0)
        \put(-105,-5){$\hat t$ }
        \put(-220,60){{{$a$}}}
    \end{picture}
    \caption{Semi-major axis as a function of the asymptotic time $\hat t$ in the non-spinning case for $\hat r=500~\hat r _{\rm ISCO}\approx 120~ \rm{AU}$. The step-like behavior is a consequence of the loss of angular momentum, which is maximized when the eccentricity oscillations are close to their maximum values (the minima of the black curve in Fig.~\ref{fig: KLspin}). }
    \label{fig: a_Schwarz}
\end{figure}
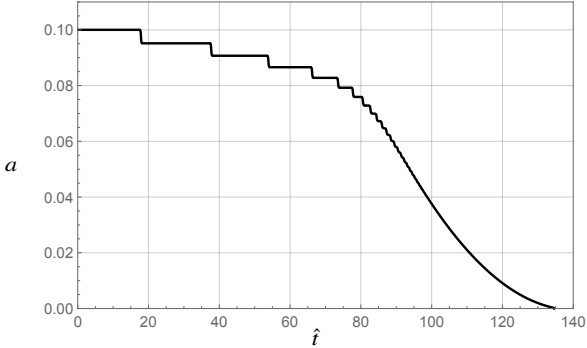

It is possible to distinguish three different phases that characterize the merger process in the example we consider. At earlier stages ($T_{\rm PN}\gg T_{\mbox{\tiny ZLK}}$) the ZLK mechanism dominates the dynamics and, due to the inner binary's high initial inclination, leads to large-amplitude oscillations for the eccentricity and inclination. More specifically one can see from Fig.~\ref{fig: KLspin} that, in agreement with the ZLK mechanism, the maximum eccentricity $e$ (minima of the black curve) corresponds to a minimum of the inclination $I$ (maxima of the cyan curve). The temporary increase of the eccentricity up to $e\approx 0.999$ leads to an efficient loss of angular momentum via GW emission in a small amount of time, which translates into the typical step-like monotonic decreasing of the semi-major axis $a$ (see Fig.~\ref{fig: a_Schwarz}), and in a progressive reduction of the ZLK oscillations amplitude.
\lbreak

In the intermediate stage, the ZLK timescale becomes comparable with the periastron precession timescale ($T_{\mbox{\tiny ZLK}}\sim T_{\rm PN}$), and the eccentricity/inclination oscillations are hampered by the PN contributions in the BBH dynamics.
\lbreak

In the final phase ($T_{\rm PN}\ll T_{\mbox{\tiny ZLK}}$), the ZLK resonance is suppressed. In other words, the BBH evolves as if the system were isolated and characterized by a high initial eccentricity inherited by the ZLK mechanism. The Peter's contributions in Eqs.~\eqref{eq:PNGWeqs} then become dominant, causing a fast orbital circularisation and a prompt decreasing in the semi-major axis of the BBH system via GW emission. Even though a realistic description of the final stages of the inspiral phase would require much more advanced analytic or numerical frameworks, here we limit to convene that the BBH merger takes place when both $a$ and $e$ vanish.

\subsection{Binary merger time in the strong field limit}
\label{sec: GW strong}
\begin{figure*}
    \centering
    \includegraphics[width=0.9\textwidth]{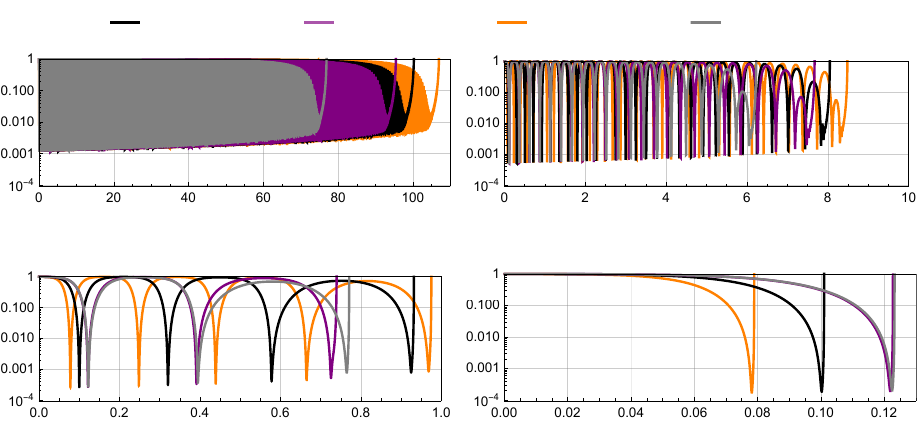}
    \begin{picture}(0,0)
        \put(-385,201){{\footnotesize{ $\chi=0$}}}
        \put(-287,201){{\footnotesize{$\chi=0.95 \left(\sigma=+1\right)$}}}
        \put(-193,201){{\footnotesize{$\chi=0.95 \left(\sigma=-1\right)$}}}
        \put(-95,200){{\footnotesize{Newtonian Case}}}
        \put(-360,-8){$\hat t$ (yrs)}
        \put(-360,95){$\hat t$ (yrs)}
        \put(-120,-8){$\hat t$ (yrs)}
        \put(-120,95){$\hat t$ (yrs)}
        \put(-285,124){$I_0=88.5^\circ$}
        \put(-285,17){$I_0=89.3^\circ$}
        \put(-60,17){$I_0=89.5^\circ$}
        \put(-50,124){$I_0=89^\circ$}
        \put(-475,135){\rotatebox{90}{{$1-e^2$}}}
        \put(-475,35){\rotatebox{90}{{$1-e^2$}}}
    \end{picture}
    \caption{The picture shows the evolution of the eccentricity for a BBH system perturbed by an external SMBH of mass $m_3=2\times 10^9 ~ M_\odot$. The four panels correspond to different values of the initial inclination $I_0$.
    The gray curve represents the case in which the external black hole is treated in the Newtonian point particle approximation, the black curve includes  general-relativistic effects associated with a non-spinning black hole whereas the purple and orange curves describe a Kerr black hole with spin parameter $\chi=0.95$ in the co-rotating (purple line) and counter-rotating (orange line) case. Here we choose $\hat r =180~ \rm{AU}$. In terms of the ISCO radius it means $\hat r \sim 1.5~\hat r_{\rm ISCO}$ for $\chi=0$ while for $\chi=0.95$ we have $\hat r \sim 4.7~\hat r_{\rm ISCO}^+$ in the co-rotating case and $\hat r \sim 1.02~\hat r_{\rm ISCO}^-$ in the counter-rotating case.
    The four panels show that small variations in the initial inclination $I_0$ correspond to huge variations in the merger time, both for the GR and the Newtonian case. The amplitude of the ZLK oscillations shows that the maximum eccentricity grows as $I_0\to90^\circ$. 
    In particular, note that when the  the ZLK mechanism is most efficient, i.e. when $I_0\sim 90^{\circ}$, 
    strong-gravity effects contribute to accelerate the BBH merger 
    compared to the Newtonian case  (bottom right panel). Instead, for smaller values of $I_0$, the ZLK mechanism is less efficient and the merger time is thus slowed down by the redshift factor (top panels and bottom left panel).    
    }
    \label{fig: KLspinstrong}
\end{figure*}
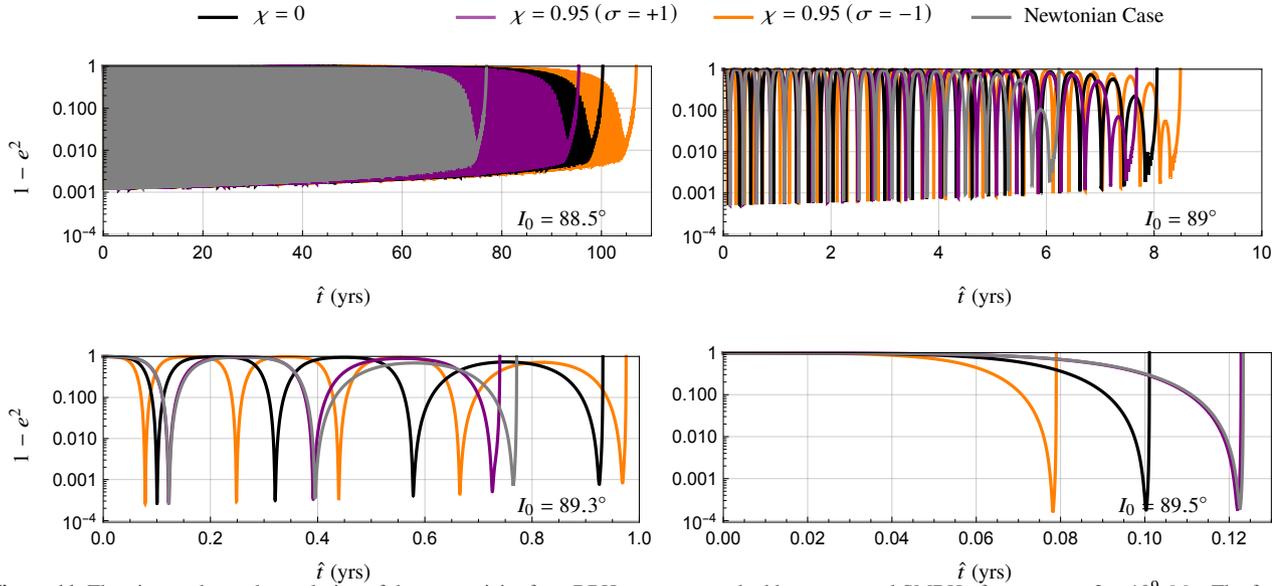  

We are now ready to study 
the evolution equations \eqref{eq:PNGWeqs} for configurations in which one can see the strong field effects in full, regarding the influence of the SMBH on the binary system. 
Indeed, we shall consider configurations for which the radius $\hat{r}$ of the outer orbit is comparable with the ISCO radius $\hat{r} _{\rm ISCO}^\sigma$ of the SMBH.
\lbreak

Since we are in a strong-gravity regime, it means that we will also be able to account for the dependence on the spin of the SMBH and in the following we will describe both the case where the BBH system moves on a co-rotating $(\sigma=+1)$ circular equatorial orbit or a counter-rotating $(\sigma=-1)$ circular equatorial orbit.
\lbreak

For the configurations of this subsection, we choose the mass of the SMBH to be $m_3=2\times 10^9~M_\odot$ which happens to be roughly the mass of M87*. This choice allows us to explore the situation where a stellar BBH system with masses $m_1=m_2=10~M_\odot$ and semi-major axis $a=0.1~\rm{AU} $ is near or at the ISCO of the SMBH, while at the same time being sufficiently far away from the tidal breakup condition \eqref{eq:breakup}, see Fig.~\ref{fig:enter-label}.
\lbreak

We consider three cases for the spin of the SMBH: the non-spinning case $\chi=0$, and the highly spinning cases $\chi=0.95$ with co-rotation ($\sigma=1$) and counter-rotation ($\sigma=-1$).%
\footnote{We choose $\chi=0.95$ since it is supposed to be close to the value of the spin of M87* as inferred by the Event Horizon Telescope Collaboration \cite{Tamburini:2019vrf, EventHorizonTelescope:2019pgp}.}
\lbreak

Finally, we choose the outer orbit radius to be $\hat r=180~\rm{AU}$, respectively corresponding to $\hat r \approx 4.68~\hat r^+_{\rm ISCO}$ for co-rotating and $\hat r \approx 1.02~\hat r^-_{\rm ISCO}$ for counter-rotating orbits. Notice that this value is also consistent in the case of a non-spinning SMBH of the same mass, since it would correspond to $\hat r \approx 1.51~\hat r_{\rm ISCO}$. We emphasize that  the chosen values of the parameters ensure that the BBH system is not subjected to a tidal breakup and can experience strong gravity effects due to the tidal interaction with the external SMBH.
\lbreak

We have solved the evolution equations \eqref{eq:PNGWeqs} for these choices of configurations. The result is depicted in 
Fig.~\ref{fig: KLspinstrong} where we  show the evolution of $1-e^2$ as a function of the asymptotic time $\hat{t}$ for four different initial inclination angles $I_0=88.5^\circ,89^\circ,89.3^\circ,89.5^\circ$. Moreover, we choose the remaining initial conditions for the inner binary to be $e_0=0.1$, $a_0=0.1~\rm{AU}$ and $\gamma_0=0^\circ$. 
As one can see, in Fig.~\ref{fig: KLspinstrong} we compare the BBH eccentricity evolution for the case where the inner binary is moving on a co-rotating and a counter-rotating circular equatorial Kerr geodesic (respectively purple and orange curves),  with the case in which the external body is a non-spinning SMBH (black curve) described in the Newtonian point particle approximation (grey curve). This is to highlight the effect of the SMBH spin.
\lbreak

We point out that when the BBH is at the ISCO of the SMBH, or close to it, the GR effects become quite significant and the Newtonian point particle approximation is no longer valid to describe the dynamics of the  system. Here we use the point particle description for the SMBH  to illustrate the comparison between our novel results valid in the full GR regime and the results obtained in the standard point particle approximation. We can conclude that describing the SMBH with the Kerr metric, instead of as a point particle, leads to  effects that significantly impact the dynamics of the BBH system. 
\lbreak

The four different panels with four different initial inclination angles  in Fig~\ref{fig: KLspinstrong} aim to illustrate the strong dependence of the merger dynamics on the initial inclination angle $I_0$: it is possible to observe that the more the BBH system is inclined, the more the ZLK mechanism is effective in reaching high values for the eccentricity and, thus, in reducing the merger time. This feature of the ZLK mechanism remains true regardless of whether the SMBH is described in a Newtonian or in a general-relativistic manner. Indeed, in all cases, one can observe that the maximum eccentricity grows as $I_0$ increases approaching $90^\circ$ and, consequently, a smaller number of ZLK oscillations is needed in order to boost the merger.
\lbreak

In particular, by making use of the terminology adopted in Ref.~\cite{Randall:2018qna}, the two top  panels and the bottom left panel in Fig.~\ref{fig: KLspinstrong} describe a \textit{slow-merger dynamics}, i.e. when the BBH system undergoes more than one ZLK cycle before merging, whereas the bottom right panel depicts a \textit{fast-merger dynamics}, in which the BBH system coalesces in just one cycle of the ZLK mechanism, and the merger time $\hat t_{\star}$ is directly given by  $\hat t_{\star}\approx u^t T_{\mbox{\tiny ZLK}}$ \cite{Randall:2018qna}.  
\lbreak

Another feature associated with strong gravity effects that emerges from Fig.~\ref{fig: KLspinstrong} is related to the impact of the redshift factor on the merger time.
We recall that in the evolution equations of the system, Eqs.~\eqref{eq:PNGWeqs}, the details of the description adopted for the SMBH enter through the ZLK frequency $\Omega_{\mbox{\tiny ZLK}}^{(\rm GR)}$ 
which now, from the point of view of an observer in the asymptotic region of the Kerr space-time, translates into the quantity $\Omega_{\mbox{\tiny ZLK}}^{(\infty)}$ defined in Eq.~\eqref{Omegainf}, where the role of the redshift factor is evident.
\lbreak

When the SMBH is described using the Kerr metric (or the Schwarzschild metric for non-spinning black holes), it is the redshift factor which is  responsible for slowing down the evolution of the system and causing an increase in the merger time compared to the case where one uses the standard Newtonian point particle description for the SMBH.  The effect of the gravitational redshift is particularly evident in the two top panels of Fig.~\ref{fig: KLspinstrong}, where the BBH evolution is characterized by a large number of ZLK cycles. Here one sees that for example the orange curve, with a redshift factor $u^t\approx 1.24 $, corresponds to a slower merger time compared to the black curve with redshift factor $u^t\approx 1.22 $ and so on.
\lbreak

Finally, Fig.~\ref{fig: KLspinstrong} highlights another consequence of being in a strong-gravity regime.
In Sec.~\ref{sec: first_look} we discussed how a general-relativistic description of the external SMBH can lead to a significant increase in the local ZLK frequency when compared to the Newtonian description. As opposed to the gravitational redshift, thus, the strong-gravity effects in the ZLK frequency contribute to catalyzing the BBH merger at earlier times than in the Newtonian case. In particular, from Fig.~\ref{fig: Omega_Ratio} it is immediate to see that for a fixed value of the radial coordinate $\hat r$ and of the spin parameter $\chi$, the deviation of $\Omega_{\mbox{\tiny ZLK}}^{(\rm GR)}$ from the Newtonian value $\Omega_{\mbox{\tiny ZLK}}^{(\rm N)}$ is greater for counter-rotating orbits.  The interplay between the GR enhancement 
of the local ZLK frequency, which accelerates the merger, and the gravitational redshift, which instead tends to slow down the merger, is particularly evident  in the bottom right panel of Fig.~\ref{fig: KLspinstrong}.  For the co-rotating case (purple), the gravitational redshift $u^t\approx 1.2$ almost entirely compensates the GR frequency enhancement $\Omega_{\mbox{\tiny ZLK}}^{(\rm GR)}/\Omega_{\mbox{\tiny ZLK}}^{(\rm N)}\approx1.21$, so that $\Omega_{\mbox{\tiny ZLK}}^{(\infty)}/\Omega_{\mbox{\tiny ZLK}}^{(\rm N)}\approx1$. This makes the purple curve almost indistinguishable from the Newtonian one (the gray curve).
In the non-spinning (black) and counter-rotating (orange) case, instead, one respectively has $\Omega_{\mbox{\tiny ZLK}}^{(\rm GR)}/\Omega_{\mbox{\tiny ZLK}}^{(\rm N)}\approx1.49$ and $\Omega_{\mbox{\tiny ZLK}}^{(\rm GR)}/\Omega_{\mbox{\tiny ZLK}}^{(\rm N)}\approx1.95$,~\footnote{This value in the counter-rotating case is consistent with the fact that the BBH orbits very close to the SMBH ISCO, $\hat r\approx 1.02~\hat r^-_{\rm ISCO}$.} corresponding to $\Omega_{\mbox{\tiny ZLK}}^{(\infty)}/\Omega_{\mbox{\tiny ZLK}}^{(\rm N)}\approx1.22$ and $\Omega_{\mbox{\tiny ZLK}}^{(\infty)}/\Omega_{\mbox{\tiny ZLK}}^{(\rm N)}\approx1.57$ respectively so that the strong-gravity effects in the ZLK frequency dominate over the gravitational redshift, thus accelerating the merger.

\subsection{GW peak frequency at the ISCO}

As already remarked, the strong-gravity effects should be maximal when the BBH is closest to the SMBH, which, for equatorial orbits, is when the outer orbit is at the ISCO. 
In this section we analyze this scenario in detail.
The scenario is realistic because interactions with the surrounding material inside the accretion disc can lead compact objects to migrate in the neighborhood of the last stable orbit and remain confined there for long enough timescales, that allow binaries to form and coalesce. For further details about these \textit{migration-trap mechanisms} we refer the reader to Ref.s~\cite{Peng_2021, Bellovary_2016, Secunda_2021}.
\begin{figure}
    \centering
    \includegraphics[width=0.40\textwidth]{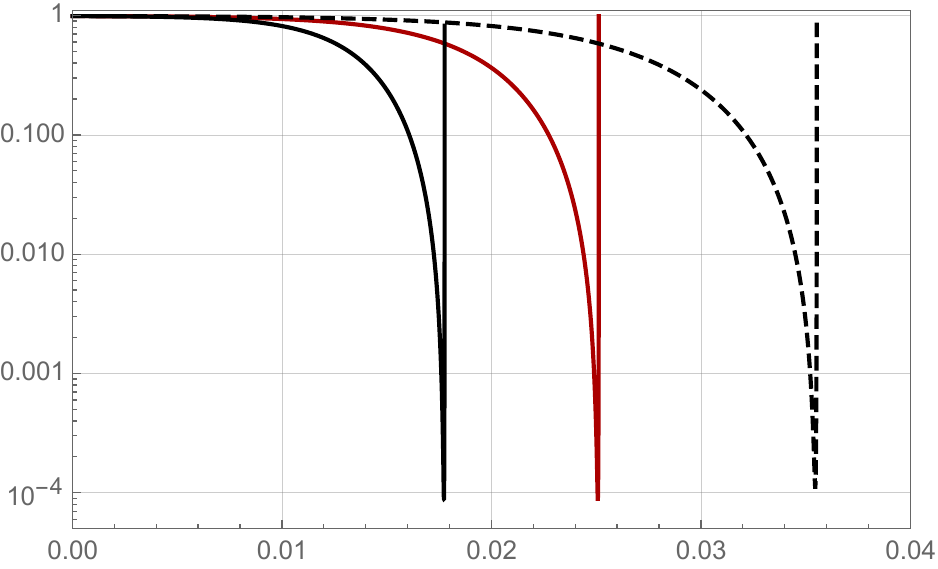}
    \begin{picture}(0,0)
        \put(-115,-8){$\tau \, / \, \hat t \hspace{1mm}\text{(yrs)}$ }
        \put(-220,50){\rotatebox{90}{{$1-e^2$}}}
    \end{picture}
    \caption{We depict here the eccentricity of a BBH system at the ISCO of an SMBH with $\chi=0$. 
    The solid black and red lines represent the system in terms of the local time $\tau$ and the asymptotic time $\hat{t}$, respectively. The masses are chosen as $m_1=m_2=10~M_{\odot}$ and $m_3=2\times 10^9~M_{\odot}$. In contrast, the dashed black line is the result obtained by using the Newtonian point particle approximation for the SMBH with the same outer orbit radius 119 AU. The initial conditions for the BBH orbital parameters are $e_0=0.1$, $a_0=0.1~ \rm{AU}$, $\gamma_0=0^{\circ}$ and $I_0=89.9^{\circ}$. }
    \label{fig: KLISCO}
\end{figure}
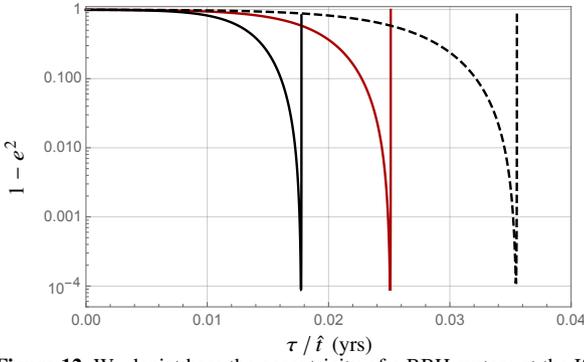
\\
\begin{figure*}
    \centering
    \includegraphics[width=0.8\textwidth]{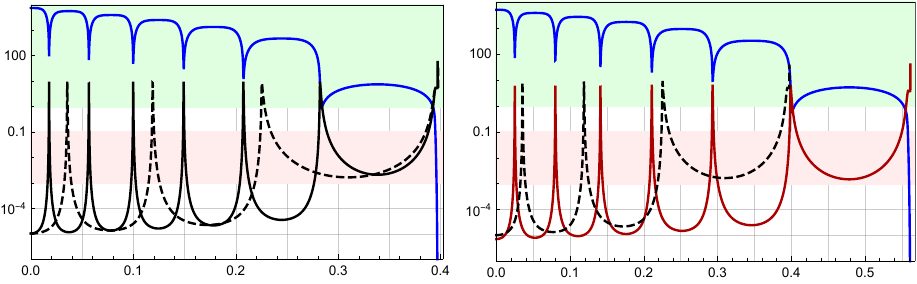}
    \begin{picture}(0,0)
        \put(-65,110){\textcolor{blue}{$T_{\text{PN}}/(u^t T_{\text{ZLK}})$ }}
         \put(-260,106){\textcolor{blue}{$T_{\text{PN}}/ T_{\text{ZLK}}$ }}
        \put(-320,-8){$\tau$ (yrs)}
        \put(-110,-8){$\hat t$ (yrs)}
        \put(-430,40){\rotatebox{90}{{$f_{\text{GW}}\left(\text{Hz}\right)$}}}
    \end{picture}
    \caption{GW peak frequency $f_{\rm GW}$  emitted by the BBH system as a function of the proper time $\tau$ (left panel) and asymptotic time $\hat t$ (right panel). The BBH is placed on the ISCO of a non-spinning SMBH with mass $m_3=2\times 10^9 ~M_{\odot}$, corresponding to $\hat r\approx 120 ~\rm{AU}$. The two black holes in the binary system have masses of $m_1=m_2=10~ M_{\odot}$ with an initial separation of $a_0=0.1~\rm{AU}$, an initial inclination angle $I_0=89.4^{\circ}$, initial eccentricity $e_0=0.1$ and $\gamma_0=0^{\circ}$.
    \\
    \textit{Left Panel:} we compare the peak frequency when treating the SMBH by a full GR description using the Schwarzschild metric (solid black) to what one obtains from a Newtonian point particle description (dashed black). One notices that there is twice the number of peaks in the GR description compared to the Newtonian approximation, consistent with the fact that $\Omega_{\mbox{\tiny ZLK}}^{(\rm GR)}=2~\Omega_{\mbox{\tiny ZLK}}^{(\rm N)}$ at the ISCO. The gravitational redshift is not considered in this plot.
    \\
    \textit{Right Panel:} the same comparison as in the left panel, but in terms of the asymptotic time $\hat t$ (red curve). This
    includes the gravitational redshift $u^t$. Notice that the inclusion of the redshift factor makes the merger time quite different in the two cases ($4.8$ months in the Newtonian case compared to $6.9$ months in the GR case) and reduces the maximum value reached by the frequency. 
    \\
    The blue curve reproduces the ratio between the PN precession and the ZLK timescales. Finally, the pink and green bands in the background respectively show the range of frequency detectable for the upcoming interferometers  ET ($1 \hspace{1mm}\text{Hz} < f_{\text{GW}}<10 \hspace{1mm}\text{kHz}$)  and LISA ($0.001 \hspace{1mm}\text{Hz} < f_{\text{GW}}<0.1 \hspace{1mm}\text{Hz}$) .}
    \label{fig: frequency}
\end{figure*}
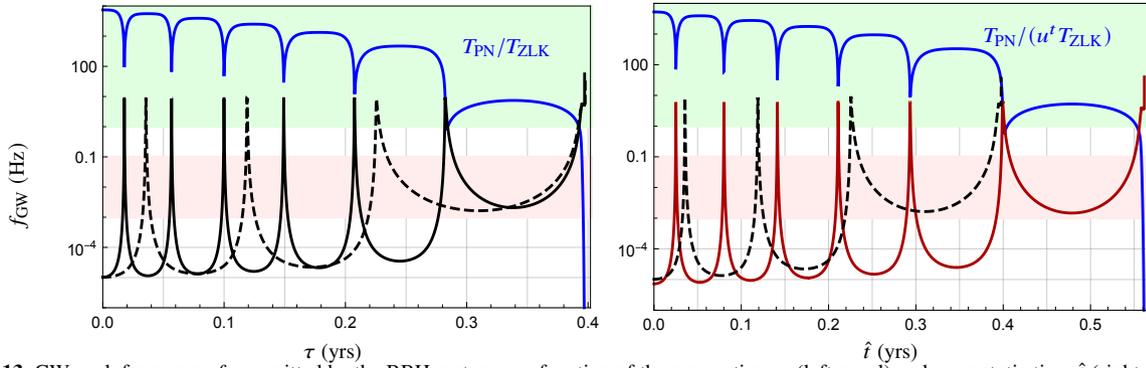
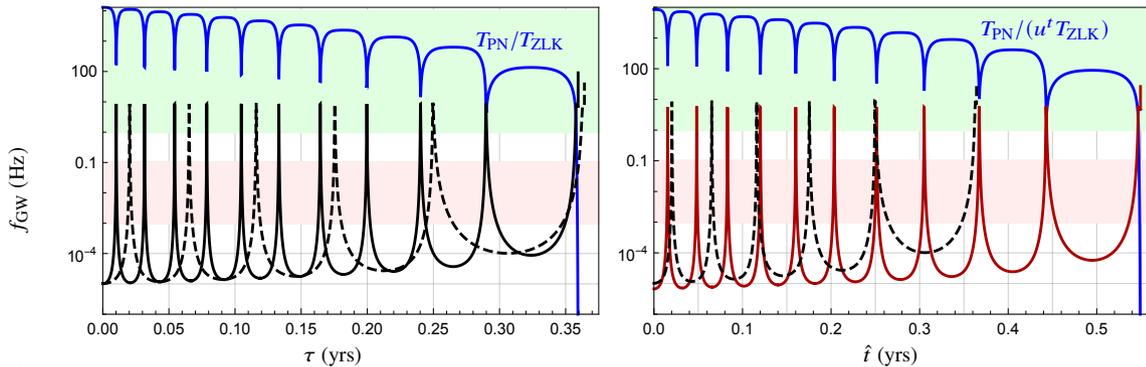
\begin{figure*}
    \centering
    \includegraphics[width=0.8\textwidth]{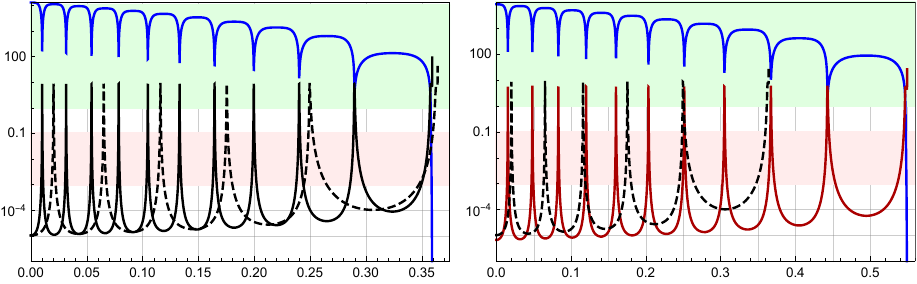}
    \begin{picture}(0,0)
        \put(-66,114){\textcolor{blue}{{$T_{\text{PN}}/(u^t T_{\text{ZLK}})$ }}}
         \put(-255,110){\textcolor{blue}{$T_{\text{PN}}/T_{\text{ZLK}}$ }}
        \put(-320,-8){$\tau$ (yrs)}
        \put(-110,-8){$\hat t$ (yrs)}
        \put(-430,40){\rotatebox{90}{{$f_{\text{GW}}\left(\text{Hz}\right)$}}}
    \end{picture}
    \caption{
    Same peak frequency comparison as in Fig.~\ref{fig: frequency} between the Newtonian and the GR case but now for a spinning SMBH with $m_3=2\times 10^9~M_{\odot}$ and $\chi=0.3$ (with $\sigma=1$). In this case, the co-rotating ISCO is located at $\hat r\approx 99 ~\rm{AU}$. As in Fig.~\ref{fig: frequency} the left and the right panels differ for the inclusion of the redshift factor, which in this case amounts to $u^t\approx 1.41$ . 
    }
    \label{fig: frequency2}
\end{figure*}

We begin with the case of a high initial inclination, $I_0=89.9^\circ$, which maximizes the ZLK mechanism. We have depicted this in Fig.~\ref{fig: KLISCO}, showing the evolution of the BBH eccentricity in terms of the proper time $\tau$ (solid black line) and the asymptotic time $\hat t $ (red line) compared with the case in which the external SMBH is treated in a Newtonian point particle approximation (dashed curve). 
The masses are chosen as in Sec.~\ref{sec: GW strong}, with $m_1=m_2=10M_\odot$ for the BBH and 
$m_3=2\times 10^9~M_\odot$ for the SMBH, ensuring that it is possible to approach the ISCO while avoiding tidal breakup. 
For simplicity we have chosen to focus on the non-spinning case $\chi=0$, for which $\hat r\approx 119~{\rm AU}$, and $u^t=\sqrt{2}$. 
\lbreak

We see from Fig.~\ref{fig: KLISCO} that the
combination of a high initial inclination angle and the strong-gravity effects at the ISCO speeds up the merger, which occurs in just a single ZLK oscillation. Accordingly, Fig.~\ref{fig: KLISCO} also shows that the maximum eccentricity is enhanced when including GR effects, reaching extremely high values $1-e^2_{\rm max}<10^{-4}$, and further contributing to reduce the merger time ($\hat t \approx 0.025~\rm{yrs}$ in terms of the asymptotic time) compared to the examples presented before in Fig.~\ref{fig: KLspin}. Finally, it is important to observe that, in terms of the proper time $\tau$ (black curve in Fig.~\ref{fig: KLspin}), the merger time is halved, as compared to the merger time when using the Newtonian point particle approximation for the SMBH (dashed curve  in Fig.~\ref{fig: KLspin}), in perfect agreement with the 
analytical prediction we derived in Sec.~\ref{sec: first_look} that at the ISCO of an SMBH one has $\Omega_{\mbox{\tiny ZLK}}^{(\rm GR)}/\Omega_{\mbox{\tiny ZLK}}^{(\rm N)}=2$.
\lbreak

Other than the merger time, it is of obvious importance to understand the frequency spectrum of the GWs emitted from the BBH, to see further observational signatures of placing the BBH at the ISCO of the SMBH. A measure of this is the peak frequency $f_{\rm GW}$ of Eq.~\eqref{eq: PeakGW} that marks the dominant frequency in the GW spectrum.
\lbreak

As pointed out in previous works \cite{Antonini:2012ad,Randall:2018qna}, an exhaustive characterization of ZLK-boosted triple systems via GW signal observations would require combining future space-based interferometers such as LISA, to probe the earlier phase of the merger when the two black holes are widely separated and the ZLK mechanism dominates, together with current ground-based facilities (such as LIGO-Virgo-KAGRA), which can detect the merger of an already-circularized BBH system at late stages.
\lbreak

As is evidenced by Figs.~\ref{fig: KLspinstrong} and \ref{fig: KLISCO}, highly inclined ($I_0>89.4^\circ$) binaries in the strong-gravity regime are characterized by fast-merger dynamics. The GW signal in these cases would result in a single pulse observed by both ET and LISA, promptly followed by a chirp signal detected by LIGO-Virgo-KAGRA. 
\lbreak

Lowering the initial inclination angle $I_0$ should instead provide more ZLK oscillations, as seen in Fig.~\ref{fig: KLspinstrong}, since the ZLK mechanism is less hampered by the periastron precession. This could in turn provide a longer signal to be observed by ET and LISA. Indeed, one would expect the detection of a BBH merger by LIGO-Virgo-KAGRA to be preceded by a series of repeated pulses detected by ET and LISA at earlier times, that mark the presence of the ZLK cycles.
\lbreak

To show this, we plot in Figs.~\ref{fig: frequency} and \ref{fig: frequency2} the GW peak frequency $f_{\rm GW}$ of the GW emission from the BBH as a function of time given a slightly lower initial inclination angle $I_0=89.4^\circ$. 
We consider two cases, one for zero spin $\chi=0$ of the SMBH (Fig.~\ref{fig: frequency}) and one with non-zero co-rotating spin $\chi=0.3$ (Fig.~\ref{fig: frequency2}). In both cases the BBH moves along the ISCO of the SMBH. In the figures we compare the results of our novel GR description for the SMBH 
to what one would have obtained by treating the SMBH as a Newtonian point particle. In detail, the black curves in Figs.~\ref{fig: frequency} and \ref{fig: frequency2} mark the peak frequency $f_{\rm GW}$ measured with proper time $\tau$ on the left while the red curve marks the asymptotically measured peak frequency $f_{\rm GW}$ with asymptotic time $\hat t$ of the Kerr space-time  on the right. These curves are then compared to the peak frequency $f_{\rm GW}$ that one obtains from SMBH as Newtonian point particle, marked by the black dashed curves.
\lbreak

While $\chi=0$ in Fig.~\ref{fig: frequency}, we have turned on the spin of the SMBH in Fig.~\ref{fig: frequency2} with $\chi=0.3$ corresponding to a co-rotating spin $\sigma=1$.
Turning on the spin parameter for the SMBH allows binary systems to access regions of space-time which would be prohibited in the non-spinning case. More specifically, the co-rotating ISCO for Kerr lies closer to the black hole than in the non-spinning case. Comparing the spinning case to the non-spinning case, we observe that the binary system undergoes a larger number of ZLK cycles before the merger occurs.
\lbreak

For the left panels of Figs.~\ref{fig: frequency} and \ref{fig: frequency2} we have depicted the evolution of the peak frequency in terms of the proper time, which is the local time of the BBH. 
These panels highlight the fact that in the early phases of the merger when the period of ZLK oscillations is much shorter than the one of the periastron precession ($T_{\rm PN}\gg T_{\mbox{\tiny ZLK}}$, as marked with the blue curve), the future interferometers ET and LISA would detect twice the number of pulses entering their frequency band (depicted with pink and green stripes respectively) compared to a Newtonian point particle description of the SMBH. This is consistent with our analytic prediction that at the ISCO of any SMBH, we have $\Omega_{\mbox{\tiny ZLK}}^{(\rm GR)}=2\Omega_{\mbox{\tiny ZLK}}^{(\rm N)}$. 
While the LISA sensitivity band ($10^{-3}~{\rm Hz}<f_{\rm GW}<10^{-1}~Hz$) detects GWs emitted only in the earlier phases of the inspiral, it is interesting to notice that the ET frequency band ($1~{\rm Hz}<f_{\rm GW}<10~{\rm kHz}$) allows to observe GWs up to the final stages of the BBH merger.
\lbreak

In the right panels of Figs.~\ref{fig: frequency} and \ref{fig: frequency2} we depict instead the evolution of the peak frequency in terms of the asymptotic time $\hat t$, {\sl i.e.}~the time appropriate for the GW detector. 
The redshift factor included in the asymptotic time is not only responsible for  reducing the maximum peak frequency reached by the GW emitted by the binary, but it also shifts the positions of the pulses and significantly affects the merger time. 
\lbreak

Finally, Figs.~\ref{fig: frequency} and \ref{fig: frequency2} show how in the last part of the merger $(T_{\rm PN}\ll T_{\mbox{\tiny ZLK}})$ the GW peak frequency enters the detectable band of LIGO/Virgo/Kagra ($f_{\rm GW}>10~\rm{ Hz}$).

\section{Conclusions}
\label{sec: concl}

In this paper, we investigated the dynamics of a BBH in point particle approximation subject to the strong gravitational field of an SMBH as described by the Kerr metric. We analyze this in the hierarchical regime where the mass of the SMBH is much bigger than the masses of the black holes in the binary system and where the size of the binary system is much smaller than the curvature scale of the SMBH. In this way, the BBH system experiences the presence of the SMBH through tidal forces which we include up to the quadrupole order.
\lbreak

The novelty of our approach is that we can allow the BBH system to move on a geodesic very close to the SMBH where strong gravity effects from the SMBH are at play.
Indeed, using our approach we can study what happens when the BBH is close to, or at, the ISCO of the SMBH.
From our analysis emerges that describing the SMBH using the Kerr
metric, instead of a point particle description, gives rise to effects that have a significant influence on the dynamics of the BBH system.
With this, we analyzed several effects which presumably could be relevant for astrophysical observations.
\lbreak

Our main results are:
\begin{itemize}
    \item We derive both a local and global description of BBH. The local description is in terms of Marck's frame of reference provided an inertial frame as the BBH moves along the equatorial circular geodesic of the SMBH. The local time near the BBH is the proper time $\tau$. 
    \item
    The global description is in terms of the distant-star frame of reference, that thanks to the stationarity and axisymmetry of the Kerr metric can be argued to have the same orientation as an asymptotic observer. In the distant-star frame, one can directly observe the gyroscope precession of the BBH in the local dynamics as a fictitious force.  Moreover, we have the asymptotic time $\hat t$ which is related to $\tau$ by the redshift factor. 
    \item We derive a secular Hamiltonian by taking the average over both the inner orbit, through the mean anomaly $\beta$, and the outer orbit. For the outer orbit one should average over the rotation angle $\hat \phi$ of the Kerr metric in the distant-star frame. In Marck's frame one should instead average over the angle $\Psi$. However, using that the gyroscope precession relates these angles, we show the equivalence of the secular dynamics via a canonical transformation. 
    \item The ZLK mechanism is modified through the ZLK frequency, and we show it is an order one effect near or at the ISCO of the SMBH. In terms of $\tau$ one gets a doubling of the ZLK frequency compared to the Newtonian point particle approach.
    \item We reproduce the dynamics previously obtained in the weak field limit, corresponding to adding 1PN terms to the Newtonian point particle approach for the SMBH.
    \item We write down the evolution equations for BBH with strong gravity influence of SMBH including periastron precession and GW radiation-reaction. 
    We see from these that the ZLK mechanism is enhanced by the strong gravity influence of the nearby Kerr black hole that describes the SMBH. We investigate also how the spin of the SMBH affects the dynamics.
    \item The enhanced ZLK mechanism provides more ZLK cycles and shorter merger times since the maximum eccentricity is larger, and thereby one has more energy emitted by GWs. At the same time, the redshift factor is seen to slow down the dynamics, since clocks run slower near the BBH as seen by an asymptotic observer.
    \item When the BBH is moving at the ISCO the asymptotic ZLK frequency is maximal for counter-rotating case, and bigger than zero spin case. Instead for the co-rotating case the asymptotic ZLK frequency is smaller with a bigger spin. 
    \item We study the peak frequency of the GWs emitted by the binary. We find that one can observe the GWs within the ET and LISA bands in the ZLK cycles before the merger, providing possible observational windows.
\end{itemize}

The analysis of this paper opens up several new directions to explore:
\begin{itemize}
    \item It would be interesting to extend our analysis to higher tidal corrections, such as the octupole contribution. This could allow for new dynamical effects in the ZLK mechanism, such as orbital flips \cite{doi:10.1146/annurev-astro-081915-023315}.
    \item One could also study the mixed terms for which the PN corrections to the BBH correction mix with the tidal forces. Here the magnetic tidal moments would play a role.
    \item It would be important to extend our analysis to other geodesics of the Kerr metric. The first case to examine could be spherical orbits outside the equatorial plane \cite{Teo:2020sey}. 
    \item It would be highly interesting to extend our analysis to extreme-mass-ratio-inspiral BBH systems, following \cite{Yang:2017aht,Camilloni:2023rra}.
    \item One could also investigate whether one can apply the EOB formalism~\cite{Buonanno:1998gg, Buonanno:2000ef}  to binary systems in the presence of tidal forces \cite{PhysRevD.89.044043}.
    \item We have investigated the long-time-scale dynamics by finding a secular Hamiltonian for the BBH. It would be very interesting to also consider the resonances in the system, which would require that one goes beyond the secular averaging approach of this paper. For papers in this direction, see~\cite{Brink:2015roa, Yang:2017aht, Gupta:2021cno, Gupta:2022fbe, Kuntz:2021hhm}.
\end{itemize}

\section*{Acknowledgments}

We thank Bin Liu, Roberto Oliveri, Andrea Placidi, and Johan Samsing for useful discussions. The authors also thank the anonymous referee for suggestions that improved the manuscript. 
F.C.,~G.G.,~M.O.,~and~D.P.~acknowledge financial support of the Ministero dell’Istruzione dell’Università e della Ricerca (MUR) through the program “Dipartimenti di Eccellenza 2018-2022” (Grant SUPER-C). G.G.~and M.O.~acknowledge financial support from Fondo Ricerca di Base 2020 (MOSAICO) and 2021 (MEGA) of the University of Perugia. 
F.C. acknowledges support by the ERC Advanced Grant “JETSET:
Launching, propagation and emission of relativistic jets from binary mergers and across
mass scales” (Grant No. 884631).


\section*{Data availability}
All data are incorporated into the article and its online supplementary
material.


\bibliographystyle{mnras}
\bibliography{_MNRAS_Bibliography} 



\bsp	
\label{lastpage}
\end{document}